\documentclass[11pt]{article}

\usepackage{amsfonts}
\usepackage{amssymb}
\usepackage{amsmath}
\usepackage{amsthm}
\usepackage{latexsym}
\usepackage{verbatim}
\usepackage{epsfig}
\usepackage{amsbsy}
\usepackage{amscd}
\usepackage{bm}

\usepackage{graphicx}

\usepackage{ifthen} 
\usepackage{xcolor}
\usepackage[colorlinks=true, linkcolor=teal, anchorcolor=teal,
            citecolor=blue,  filecolor=blue, menucolor=blue, pagecolor=teal,											urlcolor=blue, plainpages=false, pdfpagelabels]{hyperref}

\newcommand{\mymail}[1]{\href{mailto:#1}{\texttt{#1}}}

\usepackage{fancyhdr}
\usepackage[realmainfile]{currfile}
\fancypagestyle{firststyle}{

}

\newcommand{\setauthA}[1]{\def\authA{#1}}
\newcommand{\setauthB}[1]{\def\authB{#1}}

\def\printA{\begin{tabular}{l} \authA \end{tabular}}
\def\printB{\begin{tabular}{l} \authB \end{tabular}}

\newcommand{\makemytitle}[1]{\begin{center}{\textsf{\LARGE #1}}
  \end{center}
}

\usepackage[sf,bf]{titlesec}

\usepackage{algorithmic}
\usepackage{algorithm}

\usepackage[nonamebreak,square,numbers,sort]{natbib}

\usepackage[letterpaper, textheight = 676pt]{geometry}
\providecommand{\M}[1]{\mathbf#1}
\providecommand{\mc}[1]{\mathcal#1}
\providecommand{\mc}[1]{\mathcal#1}

\newcommand{\R}{{\mathbb R}}

\DeclareMathOperator{\E}{\mathbf{E}}
\DeclareMathOperator{\p}{\mathbf{P}}

\providecommand{\T}{\top} 

\DeclareMathOperator*{\argmin}{argmin}

\providecommand{\wt}[1]{\widetilde{#1}}
\providecommand{\wh}[1]{\widehat{#1}}

\providecommand{\norm}[1]{\left \lVert#1 \right \rVert}
\providecommand{\nnorm}[1]{ \lVert#1 \rVert}

\newcommand{\nscp}[2]{\langle#1, #2\rangle}

\newcommand{\dev}[2]{\Big\lvert _{{#1}={#2}}}
\newcommand{\devs}[1]{\Big\lvert _{{#1}}}

\newcommand{\blanco}[1]{  }

\newcommand{\deriv}[3]{%
\ifthenelse{#1 = 1}{\frac{d\,#2}{d\,#3}}{\frac{d^{{#1}} #2}{d{#3}^{{#1}}}}
}

\newcommand{\partials}[3]{%
\ifthenelse{#1 = 1}{\frac{\partial\,#2}{\partial\,#3}}{\frac{\partial^{#1}
    #2}{\partial#3^{#1}}}
} 

\def\su{\sum_{i=1}^n}

\def \coloneq{\mathrel{\mathop:}=}
\def \invcoloneq{=\mathrel{\mathop:}}
\def \eps{\varepsilon}

\newtheorem{theo}{Theorem}

\newtheorem{lemma}{Lemma}

\newenvironment{bew}{\begin{proof}[Proof]}{\end{proof}}

\def\R{\mathbb{R}}

\def\eps{\epsilon}

\usepackage{subfigure}
\usepackage{caption}

\newboolean{isjournal}
\setboolean{isjournal}{true}

\newboolean{nocolors}
\setboolean{nocolors}{false}

\newboolean{usemtpro}
\setboolean{usemtpro}{false}

\ifthenelse{\boolean{nocolors}}{\hypersetup{colorlinks=false}}{}

\makeatletter
\newcommand\footnoteref[1]{\protected@xdef\@thefnmark{\ref{#1}}\@footnotemark}
\makeatother

\setauthB{{\bfseries Martin Slawski}}

\setauthA{{\bfseries Niloofar Ramezani}}

\begin{document}
\thispagestyle{firststyle}

\makemytitle{{\Large {\bfseries {Lasso Penalization for High-Dimensional Beta Regression Models: Computation, Analysis, and Inference}}}}
\vskip 2.5ex
%
{\large\begin{center}
\printA
\printB
\vskip1.5ex
{\scriptsize $^{1}$Department of Biostatistics, Virginia Commonwealth University, Richmond, VA 23219, USA $\; \; \;$}\\
{\scriptsize $^{2}$Department of Statistics, University of Virginia, Charlottesville, VA 22903, USA $\; \; \;$}\\[.5ex] 
{\small \mymail{ramezanin2@vcu.edu} 
$\quad$ \mymail{ebh3ep@virginia.edu}} 
\end{center}}
\vskip 2.5ex

\begin{abstract}\vspace*{-.5ex} \noindent Beta regression is commonly employed when the outcome variable is a proportion. Since its conception, the approach has been widely used in applications spanning various scientific fields. A series of extensions have been proposed over time, several of which address variable selection and penalized estimation, e.g., with an $\ell_1$-penalty (LASSO). However, a theoretical analysis of this popular approach in the context of Beta regression with high-dimensional predictors is lacking. In this paper, we aim to close this gap. A particular challenge arises from the non-convexity of the associated negative log-likelihood, which we address by resorting to a framework for analyzing stationary points in a neighborhood of the target parameter. 
Leveraging this framework, we derive a non-asymptotic bound on the $\ell_1$-error of
such stationary points. In addition, we propose a debiasing 
approach to construct confidence intervals for the regression parameters. A proximal gradient 
algorithm is devised for optimizing the resulting penalized negative log-likelihood function. Our theoretical analysis is corroborated via simulation studies, and a real data example concerning the prediction of county-level proportions of incarceration is presented to showcase the practical utility of our methodology. 
\end{abstract}


\vspace*{-1ex}
\section{Introduction}\label{sec:intro}
\vspace*{-1.5ex}
Beta regression \cite{Ferrari2004, Simas2010, Liu2018} is a well-studied approach to regression modeling when the response variable is a proportion (e.g.,~county-level percentages of persons having health insurance, fraction of income spent on housing, proportion of land covered by forests), or more generally, contained in a bounded interval. Unlike proportions derived from binomial data, such responses are distinctly different from binomial response, which is typically modeled with a generalized linear model \cite{Mcc1989}, in that numerator and/or denominator are either unknown, non-integer, or disregarded since a Binomial model may not appear appropriate for various reasons. For example, Binomial regression assumes that variances are inversely proportional
to the number of trials so that populous counties would be treated differently from counties with
a small population. 

In this paper, we study Beta regression in a high-dimensional setup in which the number of 
predictors $p$ can exceed the sample size $n$, potentially dramatically so. Generally speaking, it is well-known that consistent estimation in such a setup can be accomplished if the underlying regression parameter is sparse, i.e., if most of the entries of the target parameter are (approximately) zero \cite[e.g.,][]{Buhlmann2011}. The use of sparsity-promoting regularization
via a suitable penalty such as the $\ell_1$-norm (aka LASSO \cite{Tibshirani1996}) has proven to 
be pivotal in this regard \cite[e.g.,][]{Candes2007, Bickel2009, Meinshausen2009, Huang2012, Greenshtein2004, vandegeer2008}. One of the primary contributions of this paper is a non-asymptotic bound on the estimation error in $\ell_1$-norm that is in line with that literature. The main challenge towards such a bound is the non-convexity of the objective function, which places the
problem outside established frameworks \cite{Negahban2012, Koltchinskii2011, Wainwright2019} relying 
on convexity. Instead, we adopt the framework developed in \cite{Elsener2018} geared towards showing 
consistency for stationary points of non-convex $\ell_1$-penalized objectives in a neighborhood 
of the true parameter. The fact that the gradient of that objective depends on the log-response, which is unbounded, requires additional care in the analysis. We note that a variety of non-convex M-estimators have been studied in 
the literature \cite[e.g.,][]{Loh2012, Loh2015, Song2020, Mei2018}. However, the body of work is scarce by comparison. The present paper adds
to this literature. 

In addition to the aforementioned $\ell_1$-bound, we also provide methodology for constructing 
confidence intervals for low-dimensional parameters of the high-dimensional target, building 
on the debiasing approach pioneered in \cite{Zhang2014} and further developed in \cite{Javanmard2014, vandegeer2014}. We present a scalable proximal gradient descent algorithm for minimizing the $\ell_1$-penalized
negative log-likelihood. We corroborate our theoretical analysis via simulation studies and conclude with an illustrative case study from criminal justice research.  

\vskip1.3ex
\noindent {\em Related work}. Model selection for Beta regression is studied in \cite{Bayer2015, Bayer2017}. Variable selection for additive models via boosting is proposed in \cite{Schmid2013}. Random forest is another widely used approach for fitting non-linear models while performing variable selection; a variant for Beta regression is developed in \cite{Weinhold2020}. Several papers discuss 
penalized estimation for Beta regression, via the ridge penalty or an $\ell_1$-penalty as considered herein \cite{Abonazel2023, Firinguetti2024, Zhao2016, Zhao2014}. The paper \cite{Abonazel2023} is dedicated to the use of a ridge penalty. The work \cite{Zhao2014} considers a double penalty
for both the means and the dispersion parameters, adopting the framework in Fan and Li \cite{Fan2001}, which includes non-convex penalties such as the smoothly clipped absolute deviation (SCAD). Accordingly, \cite{Zhao2014} present an algorithm based on coordinate descent operating on a local quadratic approximation of the penalized log-likelihood, and demonstrate the oracle property of their estimators in the spirit of \cite{Fan2001}. In particular, the analysis in \cite{Zhao2014} is limited to the low-dimensional case with a fixed number of predictors. A similar algorithm as in \cite{Zhao2014} is presented in \cite{Firinguetti2024} and evaluated via simulations. The paper \cite{Zhao2016} proposes a 
quasi-likelihood approach to Beta regression in which the objective function has the same form as that of logistic regression and the mean function is expanded in a partially linear single index model. Variable selection in this setup based on penalties as in Fan and Li \cite{Fan2001} is studied as well. 

To our knowledge, there are no prior works containing an analysis of $\ell_1$-penalized Beta regression in a high-dimensional setup. While finalizing the work on this manuscript, we became
aware of the recent work \cite{Stein2025} on sparse Beta regression models for network analysis. However, this work borrows its title from the Beta model in network analysis rather than the Beta distribution, and employs a logistic regression-style likelihood. Hence this work is not directly related to the present paper. 
\vskip1.3ex
\noindent {\em Outline}. Our notation is summarized in the paragraph below. Our approach and its 
implementation are provided in $\S$\ref{sec:methodology}. The main results of our theoretical analysis are presented in $\S$\ref{sec:analysis} and proved in the appendix. Empirical results 
in the form of simulations and the aforementioned case study are the subjects of $\S$\ref{sec:empirical}. We conclude in $\S$\ref{sec:conclusion}.  

\vskip1.3ex
\noindent {\bfseries Notation}. Frequently used symbols are summarized in Table \ref{tab:notation}. In addition, for functions $f(n)$ and $g(n)$, we write $f(n) \lesssim g(n)$ to mean that 
$f(n) \leq C g(n)$ for a universal constant $C > 0$. Equivalently, $f(n) = O(g(n))$, where $O(\cdot)$ is the Landau big-O. If $f(n) \lesssim g(n)$ and 
$g(n) \lesssim f(n)$ hold simultaneously, we write $f(n) \asymp g(n)$. We write 
$C_{\ldots}, C'_{\ldots}, C''_{\ldots}, \overline{C}_{\ldots}, \wt{C}_{\ldots}, c_{\ldots}$ etc.~to denote positive constants depending at most on the list of variables $\ldots$ in 
their subscripts. Their values may change from instance to instance.

\begin{table}[h!!!]
{\footnotesize
\begin{center}
\begin{tabular}{|ll|ll|}
\hline & & &\\[-1.5ex]
$X$ & predictor variables  & $\lambda = \lambda_n$ & regularization parameter  \\[1ex]
$Y$ & response variable  & $R_n$ &  empirical risk \\[1ex]
$\beta^*$ & ``true" regression vector   & $R$ & risk  \\[1ex]
$\phi^*$ &  ``true" scale parameter  & $\mu(z)$ & $\exp(z) / (1 + \exp(z))$   \\[1ex]
$\beta,\phi$ & generic parameter values   & $\mu(\beta)$ & $\mu(x^{\T} \beta)$ \\[1ex]
$\nabla, \nabla^2$ &  gradient, Hessian  & $\mu'(\beta) $& $\frac{d \mu(z)}{d z} \dev{z}{x^{\T} \beta}$     \\[2ex]
$\Gamma$ & Gamma function  &  $\mu_i(\beta) \, \text{etc.}$ & $\mu(X_i^{\T} \beta)$   \\[1ex]
$\Psi $ & Digamma function   & \textsf{Beta}(a,b)  & Beta distribution\\[2ex]
\hline
\end{tabular}
\end{center}}
\vspace*{-3ex}
\caption{Summary of notation used repeatedly in this paper.}\label{tab:notation}
\end{table} 

\section{Methodology}\label{sec:methodology}
\vspace*{-1.5ex}
\subsection{Approach}  
\vspace*{-.5ex}
The approach to Beta regression adopted in this paper is in line with \cite{Ferrari2004} and a corresponding 
implementation in \textsf{R}\cite{Cribari2010}. In particular, we adopt the reparameterization in terms of 
a mean parameter $\mu$ and a scale parameter $\phi$ as opposed to the parameterization of the Beta distribution that is conventionally employed outside regression contexts. For the sake of reference, the probability density function of the $\textsf{Beta}(a, b)$-distribution with parameters $a, b \geq 1$, is given by  
\begin{equation*}
f(x) = \frac{1}{\textsf{B}(a, b)} x^{a - 1} \cdot (1 - x)^{b - 1}, \quad x \in (0,1),
\end{equation*}
where $\textsf{B}(\cdot)$ denotes the Beta function. Mean and variance are given by $\mu = \frac{a}{a + b}$ and $\upsilon = \frac{a b}{(a + b)^2 (a + b + 1)}$, respectively. Defining $\phi = a + b$, the variance can be re-expressed as $\upsilon = \mu (1 - \mu) (\phi + 1)^{-1}$. Accordingly, in a regression setup we postulate that 
\begin{equation}\label{eq:beta_reg}
Y_i | X_i \sim \textsf{Beta}(\mu_i, \phi_i), \quad \mu_i = \frac{\exp(\beta_0^* + X_i^{\T} \beta^* )}{1 + \exp(\beta_0^* + X_i^{\T} \beta^*)}, \quad \phi_i \equiv \phi^*, \; 1 \leq i \leq n, 
\end{equation}
where the assumption of a homogeneous scale parameter $\phi_i^* \equiv \phi^*$, $1 \leq i \leq n$, is made for simplicity throughout this paper while acknowledging that having the scale parameter depend on covariates as well 
is a common option. 

Letting $f_i(y)$, $y \in (0,1)$, denote the PDF of $Y_i|X_i$, we have 
\begin{equation}\label{eq:cond_density}
f_i(y) = \frac{\Gamma(\phi^*)}{\Gamma(\mu_i \cdot \phi^*) \Gamma((1-\mu_i) \phi^*)} y^{\mu_i \phi^* - 1} (1 - y)^{(1 - \mu_i) \phi^* - 1}, \quad 1 \leq i \leq n,
\end{equation}
where we have used the well-known relationship $\textsf{B}(a,b) = \Gamma(a) \Gamma(b) / \Gamma(a + b)$, $a, b > 0$. 
\vskip2ex
\noindent {\bfseries Penalized Likelihood}. According to \eqref{eq:cond_density}, the negative log-likelihood for the parameters $\theta = (\beta_0, \beta, \phi)$ given $\{ (X_i, Y_i) \}_{i = 1}^n$ is obtained as 
\begin{align}\label{eq:Rn_theta}
\begin{split}
R_n(\theta) &= -\log \Gamma(\phi) + \frac{1}{n} \su \big \{ \log \Gamma(\mu_i \phi) 
+ \log \Gamma((1- \mu_i))  \\
&\qquad \qquad \qquad \qquad \qquad  - (\mu_i \phi)\log Y_i - ((1- \mu_i) \phi - 1) \log(1 - Y_i) \big \}, 
\end{split}
\end{align}
where $\mu_i = \mu_i(\beta_0, \beta) = \exp(\beta_0 + X_i^{\T} \beta)/ (1 + \exp(\beta_0 + X_i^{\T} \beta))$, $1 \leq i \leq n$. To estimate the parameters in the presence of high-dimensional predictors, we propose to minimize the $\ell_1$-penalized counterpart to \eqref{eq:Rn_theta}, i.e., 
\begin{equation}\label{eq:estimator}
\wh{\theta} \in \argmin_{\theta} \{R_n(\theta) + \lambda_n \nnorm{\beta}_1  \}, 
\end{equation}
where we note that the $\ell_1$-penalty (aka lasso penalty) excludes $\beta_0$ and $\phi$. Here, 
$\lambda_n > 0$ denotes the tuning parameter controlling the strength of the penalty. 
\vskip2ex
\noindent {\bfseries De-biasing and confidence intervals}. To construct confidence intervals for the
regression parameters, we adopt the approach pioneered in \cite{Zhang2014} and further developed in \cite{Javanmard2014} and \cite{vandegeer2014}. The rationale behind this approach is as follows. From the Karush-Kuhn-Tucker (KKT) optimality conditions associated with the minimization problem in \eqref{eq:estimator} it follows that 
\begin{equation*}
\nabla_{\beta_0, \beta} R_n(\wh{\theta}) + \lambda_n \wh{z} = 0, 
\end{equation*}
where $\nabla_{\beta_0, \beta}$ denotes the gradient operator w.r.t.~$\beta_0$ and $\beta$ and 
$\wh{z}$ is a vector whose entries are contained in $[-1,1]$. Using the Taylor approximation 
\begin{equation*}
\nabla_{\beta_0, \beta} R_n(\theta^*) \approx \nabla_{\beta_0, \beta} R_n(\wh{\theta}) + \nabla_{\beta_0, \beta}^2 R_n(\theta^*) \begin{pmatrix}
\wh{\beta}_0 - \beta_0^* \\
\wh{\beta} - \beta^*
\end{pmatrix}
\end{equation*}
and assuming that that the right hand side is close to zero (which is reasonable in light of the law of large numbers and the fact $\theta^*$ is the population minimizer), the de-biased estimator $(\wh{\beta}_0^{\text{db}}, \wh{\beta}^{\text{db}})$ is obtained as 
\begin{equation*}
\begin{pmatrix}
\wh{\beta}_0^{\text{db}} \\
\wh{\beta}^{\text{db}}
\end{pmatrix}  = \begin{pmatrix}
\wh{\beta}_0 \\
\wh{\beta} 
\end{pmatrix}  + \lambda_n \wh{\Omega} \wh{z},
\end{equation*}
where $\wh{\Omega}$ is an ``approximate inverse" of  $\nabla_{\beta_0, \beta}^2 R_n(\theta^*) \approx \nabla_{\beta_0, \beta}^2 R_n(\wh{\beta})$ satisfying 
\begin{equation}\label{eq:Omega_hat}
\nnorm{\nabla_{\beta_0, \beta}^2 R_n(\wh{\theta}) \,\wh{\Omega} - I_{d+1}}_{\infty} \lesssim \lambda_0, \qquad \lambda_0 = \nnorm{\nabla_{\beta_0, \beta}^2 R_n(\wh{\theta}) \; \{ \nabla_{\beta_0, \beta}^2 R_n(\theta^*) \}^{-1}}_{\infty},
\end{equation}
where $\lambda_0$ can be shown to be $O(\sqrt{\log(p)/n})$ in benign setups \cite{Javanmard2014}, and $\nnorm{\cdot}_{\infty}$ here denotes the entry-wise $\ell_{\infty}$-norm. Following \cite{vandegeer2014}, we expect that under suitable
conditions $(\wh{\beta}_0^{\text{db}}, \wh{\beta}^{\text{db}})$ are asymptotically normal, centered
at $(\beta_0^{*}, \beta^{*})$ and covariance matrix estimable by $\frac{1}{n} \wh{\Omega} \wh{M} \wh{\Omega}^{\T}$ with
$\wh{M} = \su \nabla_{\beta_0, \beta} R_{n}^i(\wh{\theta})^{\otimes^2}$, where $R_n^i$ constitutes
the $i$-th term in of $R_n$ in \eqref{eq:Rn_theta} and $^{\otimes^2}$ denotes the outer product of 
a vector with itself (i.e., $v^{\otimes^2} = v v^{\T}$). Standard errors and in turn confidence intervals
for the individual parameters can thus be obtained by extracting the corresponding diagonal elements of 
$\wh{\Omega} \wh{M} \wh{\Omega}^{\T} / n$. The empirical coverage of the confidence intervals obtained in this way is studied in $\S$\ref{sec:empirical}. 

\subsection{Optimization Algorithm}\label{subsec:optimization}  
To (approximately) minimize the objective function \eqref{eq:Rn_theta}, we combine the popular proximal gradient descent method \cite{Beck2009} with coordinate descent updates for the scale parameter $\phi$. Specifically, we perform a fixed number of iterations of proximal gradient descent in 
$(\beta_0, \beta)$ for fixed $\phi$, and then update $\phi$ using one-dimensional minimization for
fixed $(\beta_0, \beta)$. We alternate between these two steps until convergence. A summary is 
provided in Algorithm \ref{alg:prox}, with specific details discussed below. We note that  the objective function \eqref{eq:Rn_theta} is not globally convex, i.e., convergence to a global optimum cannot be guaranteed. However, under suitable condition (cf.~$\S$\ref{sec:analysis}), it can be shown that \eqref{eq:Rn_theta} is convex in a neighborhood of $\theta^*$, i.e., assuming that the global
minimizer and the initial iterate $\theta^{(0)} = (\beta_0^{(0)}, \beta^{(0)}, \phi^{(0)})$ are  contained in that neighborhood, global optimality can be achieved.  
\begin{algorithm}[tbh]
\caption{Optimization algorithm for minimization of \eqref{eq:Rn_theta}}\label{alg:prox}
Set $t \leftarrow 0$, initialize $\beta_0^{(0)}, \beta^{(0)}, \phi^{(0)}$. Determine stepsize $s > 0$ and fix an integer $M > 0$.  \par 
\algorithmicwhile{ {\bfseries true}}
\indent \begin{equation}\label{eq:proxgrad}
\beta_0' \leftarrow \beta_0^{(t)} - s \nabla_{\beta_0} R_n(\theta^{(t)}), \quad 
\beta' \leftarrow \textsf{S}_{\lambda \cdot s}(\beta^{(t)} - s \nabla_{\beta} R_n(\theta^{(t)})),
\end{equation}
(here, $\textsf{S}_{\tau}(z) \coloneq \big(\max\{ |z_j| - \tau, 0) \text{sign}(z_j)  \big)_{j = 1}^p$ denotes the soft shrinkage operator). \\[1ex]
\hspace*{1ex}$\qquad$ {\bfseries if} condition \eqref{eq:stepsize_cond} with $\theta' = (\beta_0', \beta')$ is not met, set 
$s \leftarrow .9 \cdot s$, and re-evaluate \eqref{eq:proxgrad} \\
\hspace*{1ex}$\qquad$ {\bfseries else}\\
\hspace*{3ex}$\qquad$ $t \leftarrow t+1$; $\beta_0^{(t)} \leftarrow \beta_0'$; $\beta^{(t)} \leftarrow \beta'$ \\
\hspace*{3ex}$\qquad$ {\bfseries if} $t \, \text{mod} \, M = 0$ \\
\hspace*{5ex}$\qquad$ $\phi^{(t)} = \argmin_{\phi} R_n(\beta_0^{(t)}, \beta^{(t)}, \phi)$\\          \hspace*{3ex}$\qquad$ {\bfseries else} \\
\hspace*{5ex}$\qquad$ $\phi^{(t)} = \phi^{(t-1)}$\\
\hspace*{3ex}$\qquad$ {\bfseries end if} \\
\hspace*{1ex}$\qquad$ {\bfseries end if}\\[1ex]
\hspace*{1ex}$\qquad$ {\bfseries if} $R_n(\theta^{(t-1)}) - R_n(\theta^{(t)}) < \textsf{tol}$ $\quad$ (here, \textsf{tol} denotes a numerical tolerance). \\
\hspace*{3ex}$\qquad$ {\bfseries break} \\
\hspace*{1ex}$\qquad$ {\bfseries end if} \\
{\bfseries end while}
\end{algorithm}

\noindent {\em Initial iterate}. The following options are considered for obtaining 
$(\beta_0^{(0)}, \beta^{(0)})$: (i) ``plain" Beta regression without penalty ($\lambda_n = 0$) using the \textsf{betareg} package \cite{Cribari2010}, or (ii) $\ell_1$-penalized logistic regression 
as implemented, e.g., in \textsf{glmnet} \cite{Friedman2010}, ignoring the fact the responses
are not in $\{0,1\}$ and that the model is misspecified. Regardless, this appears a reasonable approach particularly if (i) is not feasible (as is the case for $p > n$) because 
the associated negative log-likelihood function can still be interpreted as the cross-entropy loss 
(or ``deviance") comparing observed and fitted proportions.  

Given $(\beta_0^{(0)}, \beta^{(0)})$, we obtain $\phi^{(0)}$ via a coordinate descent step. 
\vskip1ex

\noindent {\em Choice of the stepsize}. For the proximal gradient descent method with a fixed stepsize, the default choice in the globally convex case is given by $s = 1/L$, with 
$L$ being equal to the Lipschitz constant of the gradient. Since the latter is often not 
easy to evaluate, it is common to use the upper bound $\sup_{z} \nnorm{\nabla^2 f(z)}$, where 
$\nabla^2 f(z)$ denotes the Hessian (at some point $z$) of the objective $f$ to be minimized and
$\nnorm{\cdot}$ here denotes the spectral norm (i.e., the maximum eigenvalue). Accordingly, we 
use $L = 1/ \nnorm{\nabla^2 R_n (\theta^{(0)})}$, where $\theta^{(0)}$ is the initial iterate.  In case it is computationally prohibitive to materialize $\nabla^2 R_n (\theta^{(0)})$, we use 
that $\nabla^2 R_n (\theta^{(0)}) = \frac{1}{n} \M{X}^{\T} W(\theta^{(0)}) \M{X}$ with $W = \text{diag}(w_1(\theta^{(0)}), \ldots, w_n(\theta^{(0)}))$ and the resulting 
upper bound $\nnorm{\nabla^2 R_n (\theta^{(0)}) } \leq \max_{1 \leq i \leq n} w_i(\theta^{(0)}) \nnorm{\M{X}}^2/n$, where $\M{X}$ denotes the design matrix (including intercept) and the 
$\{ w_i(\theta^{(0)}) \}_{i = 1}^n$ are observation-specific ``weights". 

A valid choice of stepsize ensures that the condition 
\begin{equation}\label{eq:stepsize_cond}
R_n(\theta^{'}) \leq R_n(\theta^{(t)}) + \nscp{\nabla_{\beta_0, \beta} R_n(\theta^{(t)})}{\theta' - \theta^{(t)}} + \frac{1}{2s} \nnorm{\theta' - \theta^{(t)}}_2^2
\end{equation}
is satisfied for all updates from $\theta^{(t)}$ to a new iterate $\theta'$, where the right hand side is used as a (local) majorization of the objective $R_n$ around $\theta^{(t)}$. If the condition \eqref{eq:stepsize_cond} fails to hold, it is reduced by a factor less than one, and a
new update is attempted until the condition is met. This procedure is in line with backtracking
line search strategies employed for proximal gradient descent methods.


\section{Analysis}\label{sec:analysis}
In this section, we state a non-asymptotic bound on the $\ell_1$-estimation error in a neighborhood of the true parameter, thereby establishing high-dimensional consistency of our approach over such neighborhood. Note that the non-convexity of the underlying objective function renders the analysis much more challenging, and as a result, existing literature has adopted a localized analysis approach. Our analysis is based on the general strategy developed in \cite{Elsener2018}.

For simplicity, we assume that the scale parameter $\phi^*$ is known. Moreover, we drop the intercept from the model, i.e., we assume that $\beta_0^* = 0$; alternatively, we may absorb the intercept in the $X$'s. With these simplifications in place, we re-state the estimation problem of interest as follows. 
\begin{equation}\label{eq:Rn_beta}
\wh{\beta} \in \argmin_{\beta} \{R_n(\beta) + \lambda_n \nnorm{\beta}_1  \}, 
\end{equation}
In the sequel, we assume that the $X_i$'s are random, and let $R(\beta) = \E[R_n(\beta)], \; \beta \in \R^p$.  
\vskip1ex
\noindent {\bfseries Assumptions}. 
\begin{itemize}
\item[{\bfseries (A1)}] The $X_i$'s, $1 \leq i \leq n$, are i.i.d.~copies of a sub-Gaussian random vector $X$ with sub-Gaussian norm \cite[][$\S$2.5]{Vershynin2018} $\sup_{v: \nnorm{v}_2 \leq 1} \nnorm{\nscp{X}{v}}_{\psi_2} < C_X < \infty$.
\item[{\bfseries (A2)}] There exists $\underline{\mu} > 0$ such that $\min_{v: \nnorm{v}_2 = 1} \{ v^{\T} \nabla^2 R(\beta^*) v \} \geq \underline{\mu}$. 
\item[{\bfseries (A3)}] \begin{equation*}
      \p(|X^{\T} \beta| \leq h) = 1 \; \text{for all $\beta \in \mc{B}$}, \; \text{where $\mc{B} = \{\beta: \; \nnorm{\beta - \beta^*}_2 \leq \eta \}$}
      \end{equation*}
      with radius $\eta > 0$ as defined in Lemma \ref{lem:SC}. In particular, with probability one, it holds that  $0 < c_h \leq \mu_i(\beta) \leq 1 - c_h < 1$, $1 \leq i \leq n$.    
\end{itemize}
The following lemma bounds the minimum eigenvalue of $\nabla^2 R(\beta)$ in a neighborhood of $\beta^*$, which yields strong convexity of $R$ in such a neighborhood. It is an important intermediate step in establishing  
our main result via Theorem \ref{theo:vdG} in the appendix. 

\begin{lemma}\label{lem:SC} Consider Assumption {\bfseries {\em (A3)}}. There exists a constant $\overline{L}_{h,\phi^*}$ depending only on $\phi^*$ and $h$ so that if $\eta = \overline{L}_{h,\phi^*}^{-1} K_X^{-1} \underline{\mu}/2$ with $K_X = 2^{3/2} C_X^{3}$ and $C_X$ as in {\bfseries {\em (A1)}}, it holds that 
\begin{equation*}
\min_{\beta \in \mc{B}} \Big\{ \min_{v: \nnorm{v}_2 = 1} v^{\T} \nabla^2 R(\beta) v \Big \} \geq \underline{\mu}/2. 
\end{equation*}
\end{lemma}
\noindent A proof of Lemma \ref{lem:SC} is provided in the appendix.

The following statement contains the main result. It establishes a high probability, non-asymptotic bound on the 
estimation error in $\ell_1$-norm, and in turn high-dimensional consistency. 
\begin{theo}\label{highdim_consistency}
Let $\lambda_n > \lambda_{\eps}$ with 
\begin{align*}
\lambda_{\eps} &= \phi^* \left\{ 2\log(2n) \left[  \sqrt{\frac{\log p}{n}} + \frac{8h }{3} \frac{\log p}{n}  
\right] +  C_{h, \phi^*} C_X \sqrt{\frac{\log p}{n}} + \wt{C}_{h,\phi^*} \frac{\log p}{n}  \right\}  + \\
&\quad+ C_X \overline{C}_{h,\phi^*} \left( \sqrt{\frac{\log(p+1)}{n}} + \frac{\log(p+1)}{n} \right). 
\end{align*}
Under Assumptions {\bfseries {\em (A1)}}, {\bfseries {\em (A2)}}, and {\bfseries {\em (A3)}}, for any stationary point $\wh{\beta}$ of \eqref{eq:Rn_beta} contained in $\mc{B}$ it holds that 
\begin{equation*}
\nnorm{\wh{\beta} - \beta^*}_1 \leq \frac{2}{\underline{\mu}} \frac{\{ \frac{3}{2} \lambda + \frac{1}{2} \lambda_{\eps} \}^2}{\lambda - \lambda_{\eps}} s,  \qquad s \coloneq\sum_{j = 1}^p \M{1}_{\beta_j^* \neq 0},
\end{equation*}
with probability at least $1 - C_{h,\phi^*}/n - \log_2(\eta n \sqrt{p})/p$, where 
$C_{h,\phi^*}, \wt{C}_{h,\phi^*}$ and $\overline{C}_{h, \phi^*}$ are positive constants depending only on $h$ and $\phi^*$.   
\end{theo}
\noindent To better understand the implications of the above theorem, let us consider the choice 
$\lambda_n \asymp \lambda_{\eps} \asymp \log(n) \sqrt{\log(p)/n}$ under the mild assumption that $n \geq \log p$, in which case $\nnorm{\wh{\beta} - \beta^*}_1 = O(s \lambda_{\eps}) = O(s \log(n) \sqrt{\log(p)/n})$ with $s$ being the sparsity level, i.e., the number of non-zero coefficients in $\beta^*$. This yields consistency under the condition $s^2 \log^2(n) \log(p)/n \rightarrow 0$ as $n \rightarrow \infty$, and roughly aligns with existing results for other high-dimensional generalized linear models \cite{vandegeer2008, vandegeer2014, Wainwright2019}. A notable difference 
is the presence of the extra $\log^2(n)$ factor, which in the proof arises from the sub-exponential tails 
of log-Beta random variables. It remains unclear whether this term can be avoided. The specific constants appearing in the theorem and their significance are provided in the appendix, which contains a detailed proof of the result. 

We would like to point out that in accordance with the framework in \cite{Elsener2018}, Theorem \ref{highdim_consistency} makes a statement about stationary points of the penalized negative log-likelihood $R_n$ in a neighborhood of $\beta^*$. Note that in practice, a stationary point generated by the proximal gradient descent algorithm proposed in $\S$\ref{subsec:optimization} is not guaranteed to be contained in this neighborhood. Establishing such a result is not attempted herein but we point out that analyses of this flavor for other $\ell_1$-penalized problems with non-convex objective exist in the literature \citep[e.g.,][]{Loh2015, Song2020, Loh2012}.       

\section{Empirical results}\label{sec:empirical}
In this section, we present the results of simulation studies to back up the theoretical  
analysis in $\S$\ref{sec:analysis} and demonstrate the utility of the approach through a case study from social justice research. 

\subsection{Simulations}\label{subsec:simulations}

{\em Setup}. Our simulations studies are broadly divided into (i) a low-dimensional setup with $s \in \{2,5,10\}$
and $p \in \{20, 50, 100\}$, and (ii) a high-dimensional setup with $s \in \{5, 10, 20\}$ and 
$p \in \{200, 500, 800 \}$. In both settings, we let $n \in \{500, 1000\}$. Here, $s$ denotes 
the size of the support of the target $\beta^*$, which is generated as follows. 
\begin{equation}\label{eq:beta_star_gen}
\beta_{2j - 1}^* = \sqrt{1/(j+1)}, \; \beta_{2j}^* = -\sqrt{1/(j+1)}, \;\; j=1,\ldots,\lceil s/2 \rceil,\; \beta_{j} = 0 \; \forall j \geq s+1, 
\end{equation}
Subsequently, $\beta^*$ is normalized to have squared Euclidean norm equal to $13/6$\footnote{This number is the squared Euclidean norm of the vector obtained according to \eqref{eq:beta_star_gen} when setting $s = 6$.}. Furthermore, we set $\beta_0^*$ and $\phi^* = 4$. The covariates $X_i = (X_{i1},\ldots,X_{ip}{})^{\T}$ are i.i.d.~Gaussian random vectors with zero mean and identity covariance, $1 \leq i \leq n$. Conditional on $X_i$, the response $Y_i$ is drawn from the Beta distribution in Eq.~\eqref{eq:beta_reg}, $1 \leq i \leq n$.

Given the data $\{ (X_i, Y_i) \}_{i =1}^n$, we consider the optimization problem \eqref{eq:estimator} with the choice $\lambda = 0.2\sqrt{\log(p)/n}$, where the rate $\sqrt{\log(p)/n}$ is the default
in $\ell_1$-penalized regression problems and the specific factor $0.2$ was determined by experimenting with values from the grid $\{0.1, 0.2, \ldots, 1\}$ in trial runs. The proximal gradient algorithm in $\S$\ref{subsec:optimization} is used to compute $(\wh{\beta}_0, \wh{\beta}, \wh{\phi})$, and confidence intervals (CIs) via de-biasing are obtained by computing $\wh{\Omega}$ in \eqref{eq:Omega_hat} with the code \cite{sslasso} and the choice $\lambda_0 = 0.01\sqrt{\log(p)/n}$. CIs are only computed for the low-dimensional setup (i). In the high-dimensional setup, coverage tends to be below
the nominal level for most $(p,s,n)$-configurations, which does not come as a surprise since correctness of the CIs is tied too more stringent conditions than consistent parameter estimation even for plain linear models with sub-Gaussian errors \cite{Javanmard2014, vandegeer2014}. Every $(p,s,n)$ configuration is run with 1,000 independent replications.  
\begin{figure}[t]
\centering
\includegraphics[width = .32\textwidth]{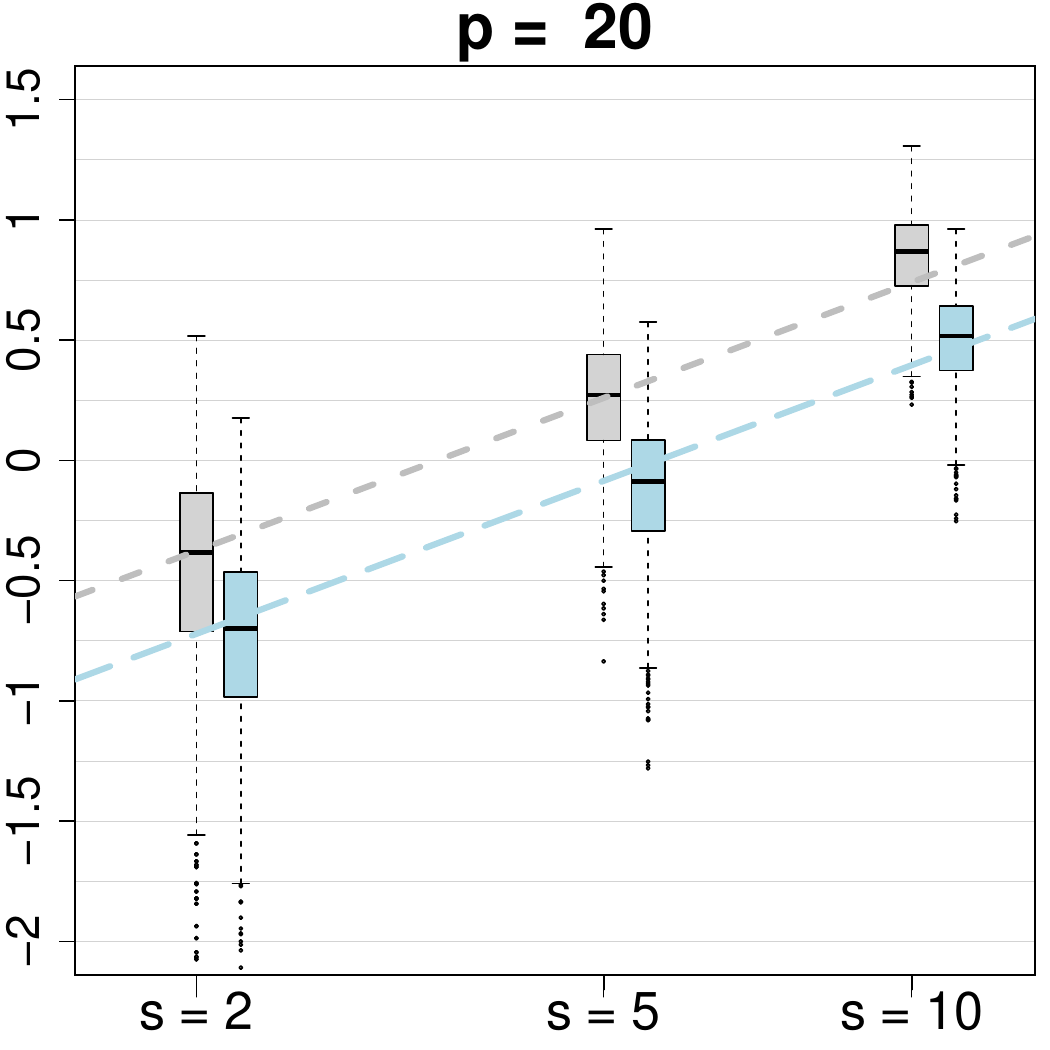}
\includegraphics[width = .32\textwidth]{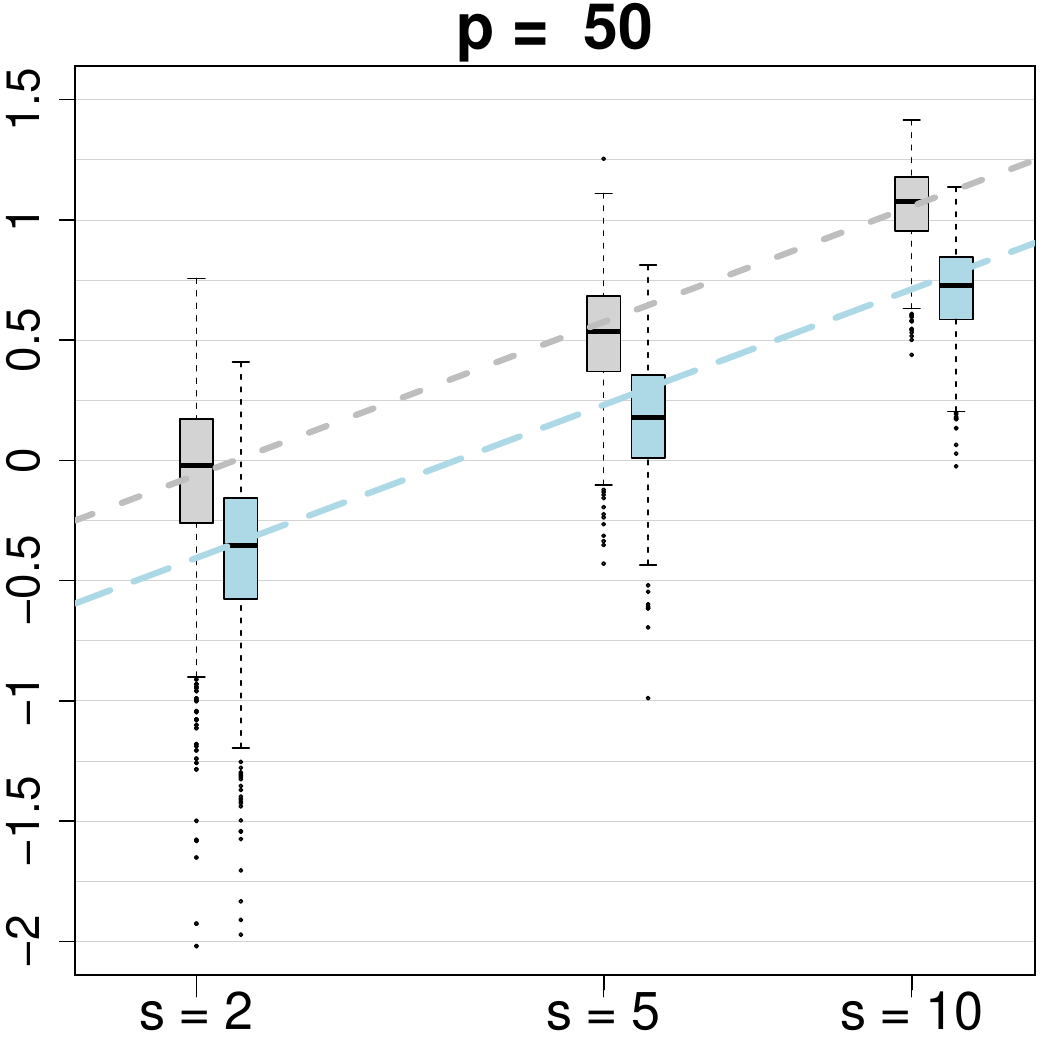}
\includegraphics[width = .32\textwidth]{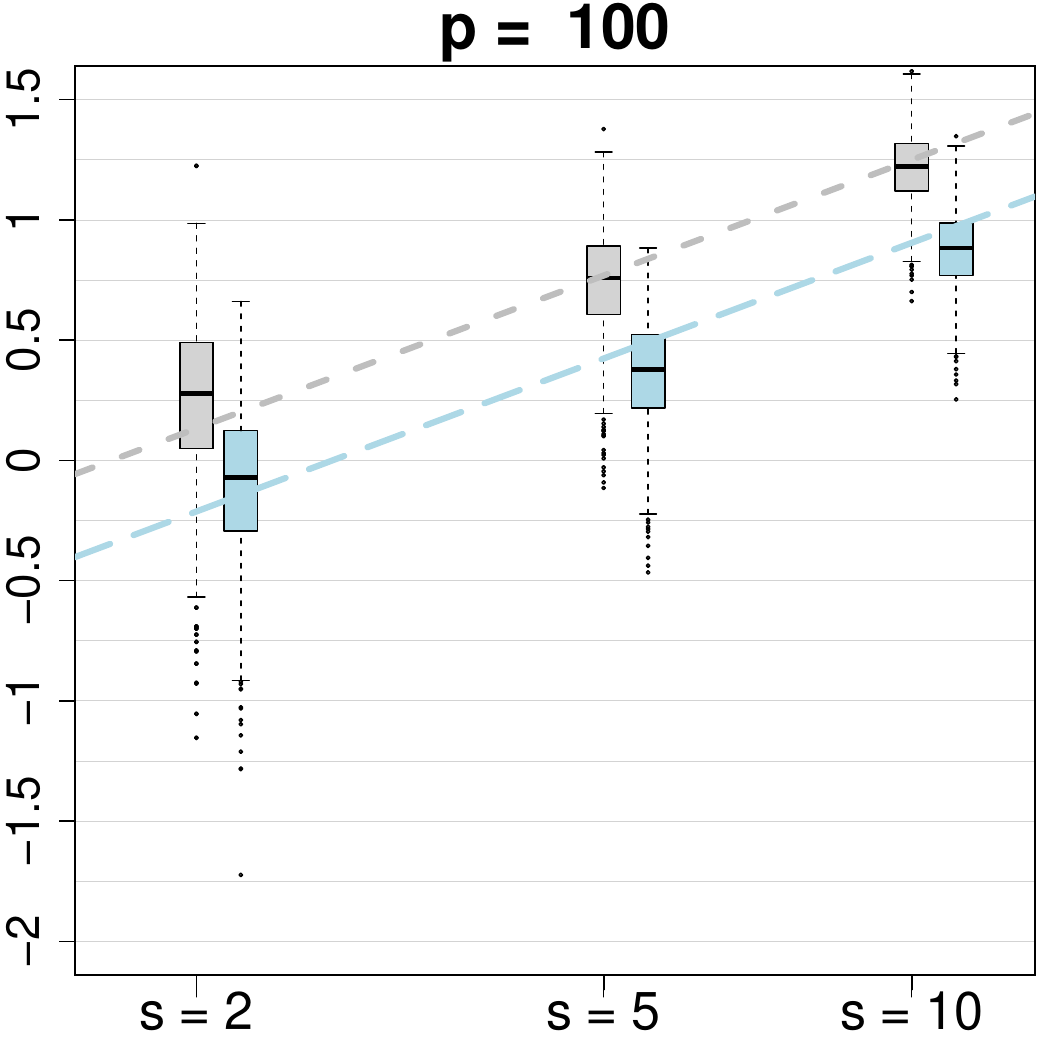}
{\footnotesize Regression model:  $\log(\nnorm{\wh{\beta} - \beta^*}_1) = \gamma_0 + \gamma_1 \log s + \gamma_2 \log n + \gamma_3 \log \log p$. \\[1.5ex]
\begin{tabular}{|l|l|l|l|l|l|} \hline
& $\wh{\gamma}_0$ & $\wh{\gamma}_1$ & $\wh{\gamma}_2$ & $\wh{\gamma}_3$ & $R^2$ \\[.5ex]
\hline
\text{low-dim} & .937 (.044) & .694 (.003) & $-$.498 (.006) & 1.18 (.012) & .77  \\
\text{high-dim} & $-.270$ (.034) & .502 (.002)  & $-$.516 (.004). & 2.21 (.013) & .85  \\
\hline
\end{tabular}}\\[1.5ex]
\includegraphics[width = .32\textwidth]{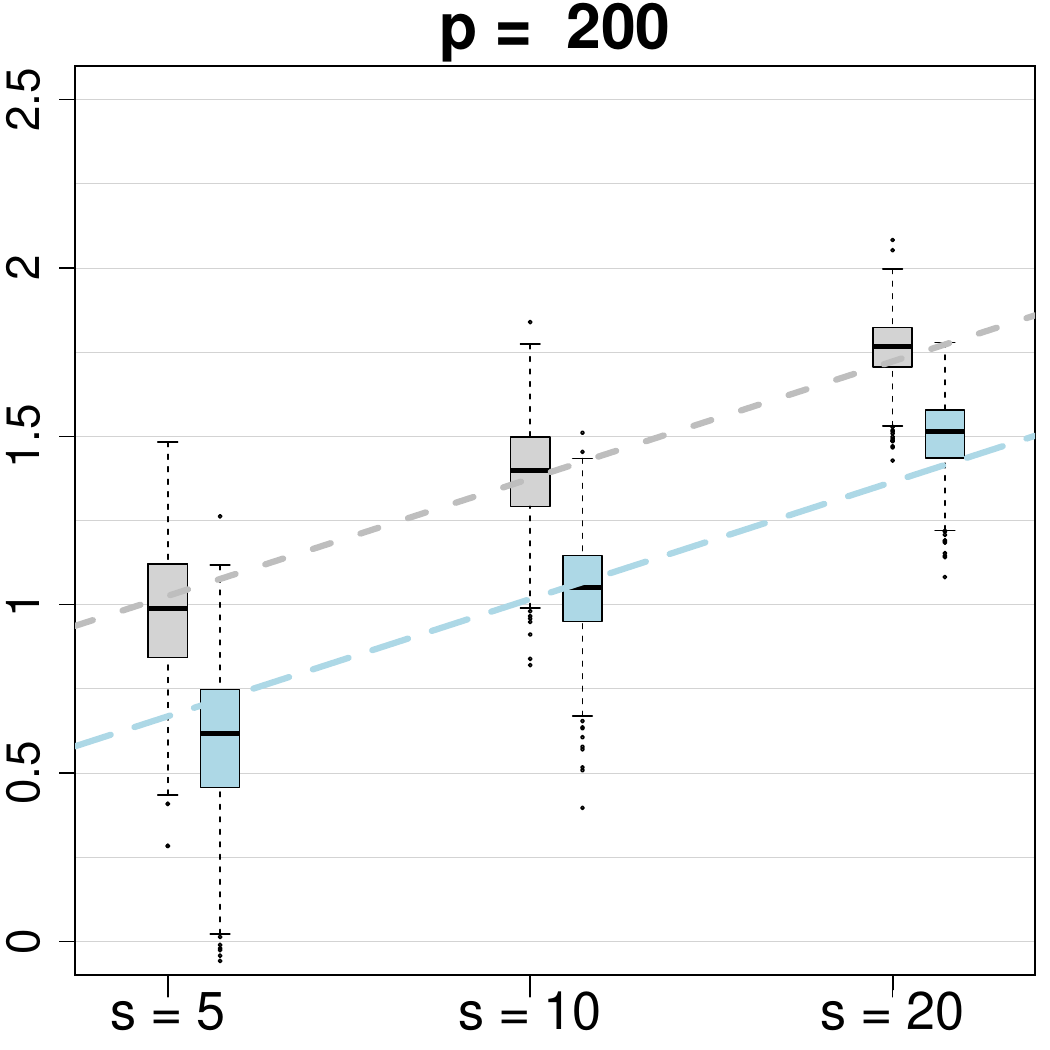}
\includegraphics[width = .32\textwidth]{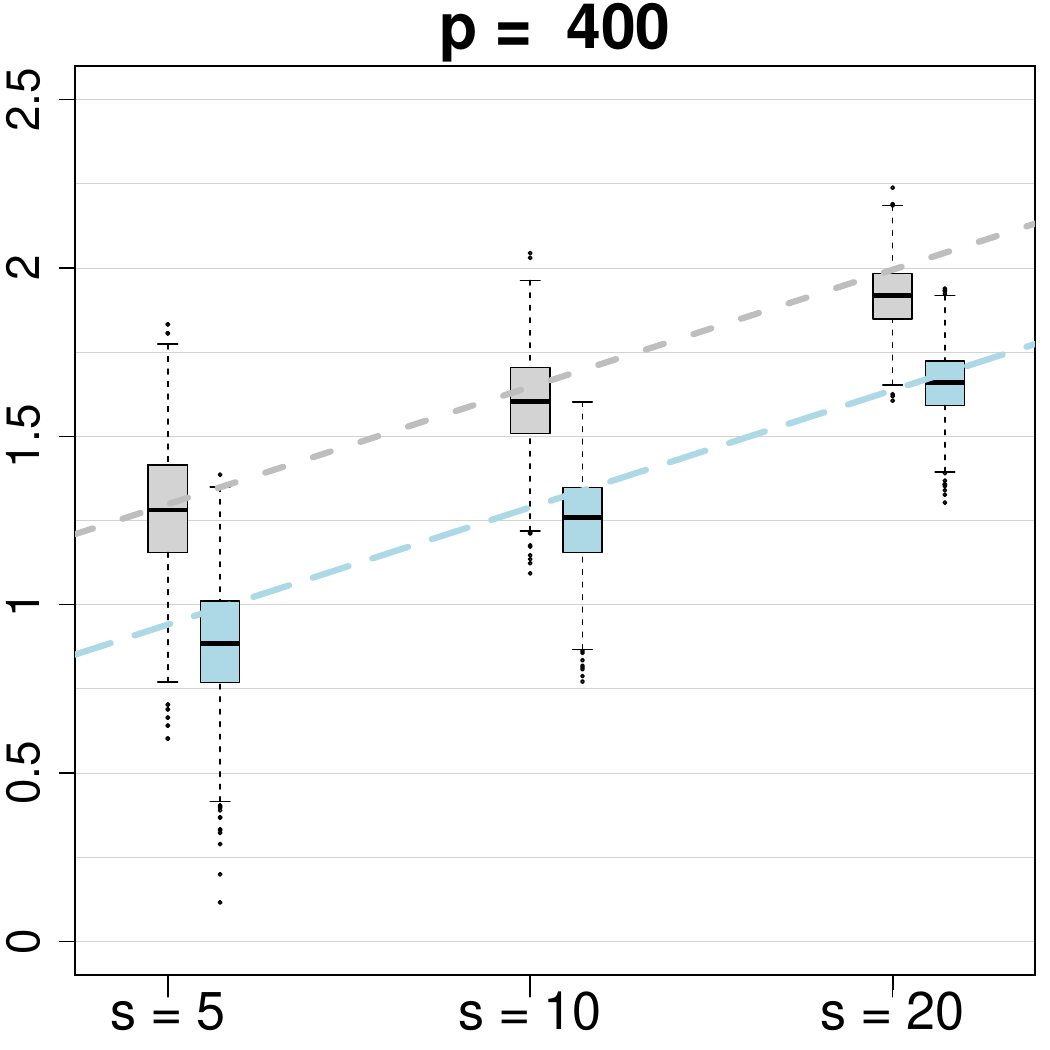}
\includegraphics[width = .32\textwidth]{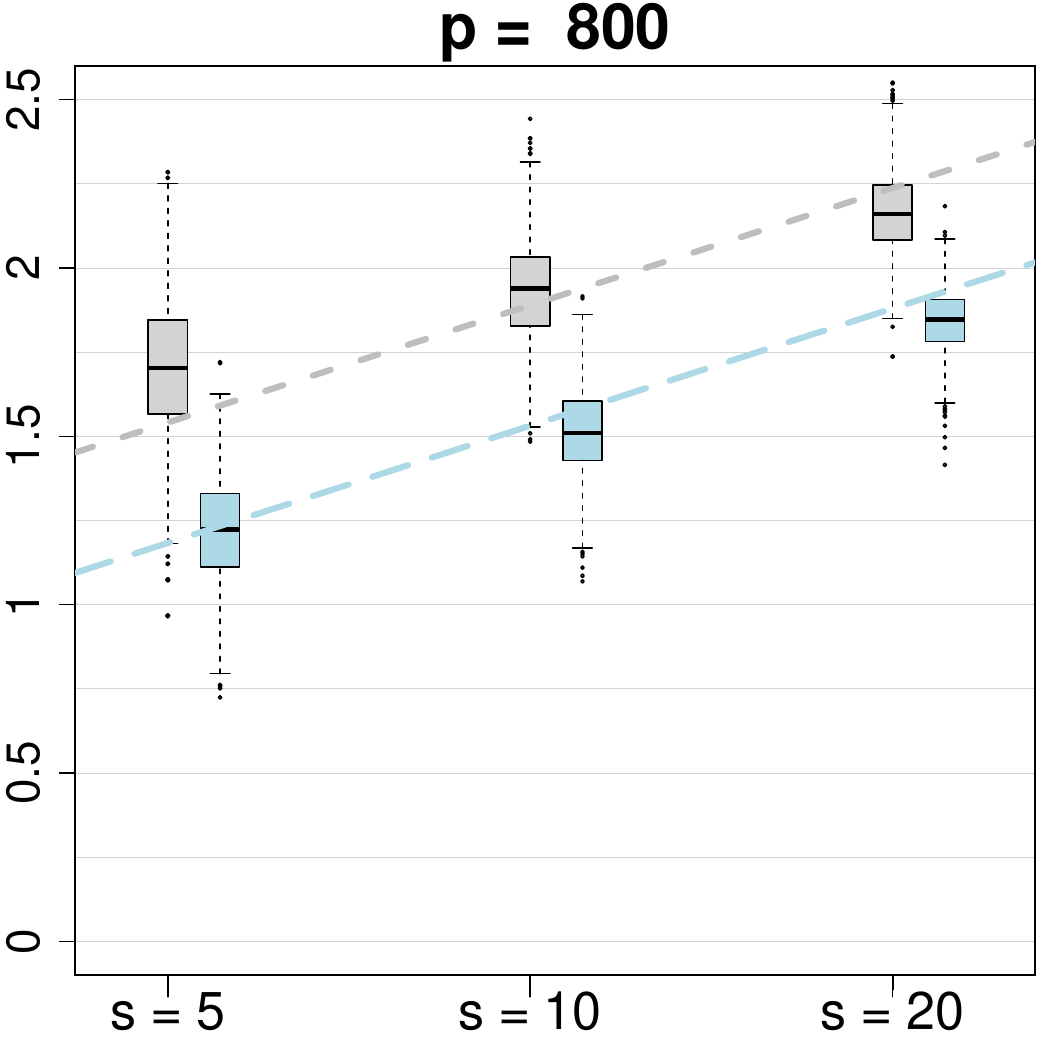}
\vspace*{-1ex}
\caption{Logarithmized $\ell_1$-estimation error $\nnorm{\wh{\beta} - \beta^*}_1$ against $s$ (axis marks on the horizontal axis also follow a log-scale), for different $(n,p)$ combinations. Grey: $n = 500$; Blue: $n = 1,000$. The table in the center contains the estimated regression coefficients and the $R^2$s of the linear model \eqref{eq:linreg} fitted for the low-dimensional (top) and the high-dimensional setup (bottom). }\label{fig:simulations}
\end{figure}
\vskip1.5ex
\noindent {\em Results}. The results of our simulations are summarized in Figure \ref{fig:simulations} and Table \ref{tab:simulations}. The figures show boxplots of the $\ell_1$-estimation errors $\nnorm{\wh{\beta} - \beta^*}_1$ in dependence of the sparsity levels $s$ for different combinations of $(n,p)$ on a log-log scale. The empirical errors are compared to predictions
according to the linear model 
\begin{equation}\label{eq:linreg}
\log \nnorm{\wh{\beta} - \beta^*}_1 = \gamma_0 + \gamma_1 \log s + \gamma_2 \log n + \gamma_3 \log \log p.
\end{equation}
Based on Theorem \ref{highdim_consistency} we expect that with a suitable choice of $\lambda$, we have
$\gamma_1 = 1$, $\gamma = -1/2$, and $\gamma_3 \geq 1/2$. A linear regression model of the form \eqref{eq:linreg} is fitted for both the low-dimensional and high-dimensional setup based on $3 \times 3 \times 2 \times 1,000 = 18,000$ simulation runs in each case. The above linear model
achieves an $R^2$ of $.77$ and $.85$ in the low-dimensional and high-dimensional setup, respectively. The estimates for $\gamma_2$ are close to $-1/2$ in both cases as expected, while the estimates 
for $\gamma_1$ are below the expected value of $1$. The estimates for $\gamma_3$ are around $1$ and $2$, respectively, confirming that the power in the polylog term in front of the $s/\sqrt{n}$ rate for the $\ell_1$-estimation error is likely larger than $0.5$, in alignment with Theorem \ref{highdim_consistency}.  

Table \ref{tab:simulations} contains average CI coverage rates (for the low-dimensional setup, in \%) and true/false positive rates (TPRs and FPRs, in \%) to evaluate the level of model parsimony achieved by means of $\ell_1$-penalization, with true and false positives corresponding to the non-zero entries in $(\wh{\beta}_j)_{j = 1}^s$ and $(\wh{\beta}_j)_{j = s+1}^p$, respectively. CI coverage rates are rather close to the nominal 95\% for $n = 1,000$ and 
$p = 20$, which roughly corresponds to the regime encountered in the case study below. Coverage rates
drop gradually (but not dramatically) as $p$ and/or $s$ are increased. For $n = 500$, the rates 
range from about 94\% to 92.5\% for $p \leq 50$ and $s \leq 5$, but to drop 90.2\% for $s = 10$. As mentioned above, it is well-known that the success of the de-biasing approach for constructing 
CIs for $\ell_1$-penalized estimators relies on the condition $s^2 \lesssim n$. Regarding TPRs and 
FPRs, we observe that the FPRs are consistently low and thus parsimonious model fits are generated. At the same time, the FPRs are also bounded away from zero, which means that selected models tend to include extraneous variables in addition to the support of $\beta^*$. This is well in line with the literature \cite{Zhao2006, Su2017}. TPRs are at least in the nineties for $s \leq 10$. Note that $s = 20$, the coefficients in the support of $\beta^*$ tend to be smaller in magnitude according to the protocol described after \eqref{eq:beta_star_gen}, increasing the chance of false non-selections due to the shrinkage induced by the $\ell_1$-penalty. 

\begin{table}
\begin{center}
{\footnotesize \begin{tabular}{|ll|ccc|ccc|}
\hline
   &    & \multicolumn{3}{c|}{$n = 500$} &\multicolumn{3}{c|}{$n = 1000$} \\
         \cline{3-5} \cline{6-8}
    &     & \multicolumn{3}{|c|}{$p$} & \multicolumn{3}{c|}{$p$} \\  
   \cline{3-5} \cline{6-8}
    &     & $20$ & $50$ & $100$  & $20$ & $50$ & $100$ \\  
    \hline
        & \vline $\,$ CVG & 93.8 (.12)& 92.8 (.17) & 92.3 (.17) &  95.1 (.06)         & 94.3 (.12) & 93.5 (.14) \\
 $s = 2$ &\vline $\,$ TPR & 100 (.00)& 100 (.00) & 100 (.00) & 100 (.00)   &  100 (.00) & 100 (.00)                 \\
         &\vline $\,$ FPR & 19.0 (.29) & 13.9 (.16) & 11.3 (.11) &    18.5 (.30)      & 13.1 (.15) & 10.4 (.10)         \\
         \hline
         & \vline $\,$ CVG & 93.7 (.11) & 92.5 (.17) & 91.7 (.22)&   95.0 (.06)  &  93.7 (.12) & 93.3 (.14)          \\
 $s = 5$ & \vline $\,$ TPR & 99.4 (.10) & 99.1 (.13) & 98.5 (.17) & 100 (.00) & 100 (.00)  & 100 (.00)                     \\
         & \vline $\,$ FPR & 19.3 (.32)& 14.1 (.17) & 11.4 (.11) & 18.8 (.32) & 13.5 (.17) & 10.4 (.10)            \\
         \hline
         & \vline $\,$ CVG  & 90.3 (.20) & 90.2 (.24) & 90.5 (.25) & 94.7 (.06) & 93.4 (.12)  & 93.2 (.14)           \\
 $s = 10$ & \vline $\,$ TPR & 88.6 (.31)  & 84.6 (.35) & 81.2 (.37) & 98.8 (.10) & 98.0 (.14) & 97.1 (.17)                     \\
          & \vline $\,$ FPR & 18.1 (.39) & 14.2 (.18) & 11.2 (.11) & 18.7 (.40) & 13.2 (.17)  & 10.3 (.10)    \\
          \hline
\end{tabular}
\vskip2ex
\begin{tabular}{|ll|ccc|ccc|}
\hline
   &    & \multicolumn{3}{c|}{$n = 500$} &\multicolumn{3}{c|}{$n = 1000$} \\
         \cline{3-5} \cline{6-8}
    &     & \multicolumn{3}{|c|}{$p$} & \multicolumn{3}{c|}{$p$} \\  
   \cline{3-5} \cline{6-8}
    &     & $200$ & $400$ & $800$  & $200$ & $400$ & $800$ \\  
    \hline
 $s = 5$ &\vline $\,$ TPR & 97.8 (.21) & 97.4 (.22)  & 97.1 (.23) &  100 (.00)  & 100 (.03)  & 100 (.00)                  \\
         &\vline $\,$ FPR & 9.0 (.07)  & 7.4 (.05) & 6.8 (.05)  & 8.2 (.07)    & 6.6 (.04) &  5.5 (.03)      \\
         \hline
 $s = 10$ & \vline $\,$ TPR & 78.0 (.39) &  74.8 (.42) & 71.7 (.42) & 96.5 (.18) & 95.4 (.21)   & 94.8 (.22)                     \\
         & \vline $\,$ FPR & 8.9 (.07) & 7.3 (.05) & 6.6 (.05)  & 8.2 (.06)  & 6.6 (.04)  & 5.5 (.03)             \\
         \hline
 $s = 20$ & \vline $\,$ TPR & 52.5 (.32) & 49.1 (.32) & 46.1 (.32) & 78.2 (.27) & 74.9 (.29) & 72.7 (.30)                      \\
          & \vline $\,$ FPR & 8.9 (.07) & 7.2 (.05) & 6.5 (.04)  &  8.1 (.07) & 6.5 (.04) & 5.4 (.03)    \\
          \hline
\end{tabular}
}
\end{center}
\vspace*{-2.5ex}
\caption{Average coverage rates (CVG) and TPRs/FPRs (in \%) with respect to the support of $\beta^*$ for various configurations of $(p,s,n)$.}\label{tab:simulations}
\end{table}

\subsection{Case Study}\label{subsec:case_study}
Our case study uses data from \cite{ramezani2022} in which macro-county characteristics predictive of jail population size were identified for all 3,141 counties in the U.S. In this study, a beta regression model was fitted using a pre-selected set of variables to understand how demographic, socioeconomic, and variables pertaining to physical \& mental health and the criminal legal system relate to jail population per capita (the target variable) at the county level. The incarceration trends database of the Vera institute \cite{Vera2018} was used to obtain incarceration counts.  
A variety of economic, social, demographic and public health-related variables were extracted
from the Robert Wood Johnson Foundation’s County Health Rankings and Roadmaps, which itself is compiled from several data sources provided by the University of Wisconsin \cite{Wisconsin2018}.
Police data were obtained from the Uniform Crime Report \cite{DOJ_2011a, DOJ_2011b}. The above data sources were linked using county and state identifiers, and per capita rates
of the aforementioned variables were calculated using the population counts from the U.S.~Census \cite{Census_Bureau_2020}. The list of variables under consideration are tabulated in Table \ref{tab:variables_casestudy}.
\begin{table}
\begin{center}
{\small \begin{tabular}{|l|ll|}
 \hline
  & & \\[-1.5ex]
Income inequality ratio  &  &Community Mental Health Centers$^{\star}$ \\               
High School graduation rate  &  &Violent crimes$^{\star}$  \\         
County size is medium (indicator) &  &Police officers$^{\star}$ \\                                  
County size is large (indicator)  & &Median household income \\                              
Poor physical health days  & & \% African American \\   
Poor mental health days & & \% Hispanic \\         
 Primary Care Physicians$^{\star}$ & &\% Rural population \\                          
 Healthcare costs$^{\star}$ & & \% Over Age of 18  \\                              
 \% drug treatment paid by Medicaid & &  \% Children in poverty \\           
 Psychiatrists$^{\star}$     &  & HPSA Score$^{\ast}$   \\[1ex] 
 \hline
\end{tabular}}
\end{center}
\vspace*{-3ex}
\caption{List of predictor variables used in the case study. Starred variables are per capita.\\ 
$\ast$: Health Professional Shortage Area score.}\label{tab:variables_casestudy}
\end{table}
\vskip1ex
\noindent {\em Results}. The $p = 20$ variables are used as predictor variables for the target 
variable ``proportion of individuals incarcerated". After eliminating counties for which 
any of the predictor variables is missing, we are left with $n = 2377$ observations. Prior to fitting our proposed $\ell_1$-penalized Beta regression approach, all predictor variables are centered and scaled to unit empirical variance as customary in penalized regression. For ease 
of presentation, results pertaining to coefficient estimates reported below are with respect to 
the modified location and scales obtained through centering and scaling. 

\begin{figure}
\begin{center}
\includegraphics[width = 0.7\textwidth]{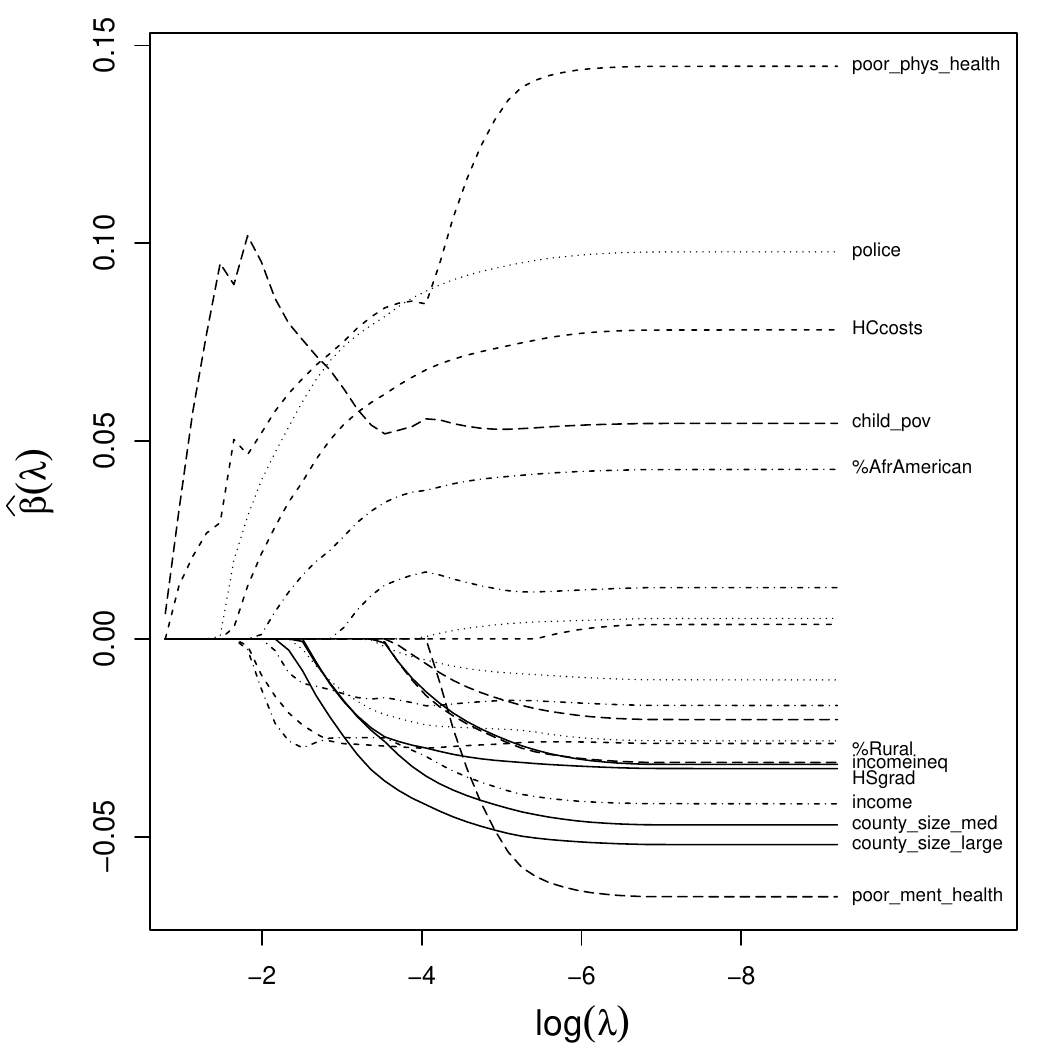}
\end{center}
\vspace*{-4.5ex}
\caption{Coefficient trajectories $\{ \wh{\beta}_j(\lambda) \}_{j = 1}^p$ for the case study. Abbreviations of the variables are given at the right.}\label{fig:solutionpath}
\end{figure}

Figure \ref{fig:solutionpath} displays the trajectories of the coefficients $\{ \wh{\beta}_j(\lambda) \}_{j = 1}^p$ for 50 values of $\lambda$ spaced evenly (on a logarithmic scale) between $10^{-4}$ and $.95 \overline{\lambda}$ with $\overline{\lambda} = \nnorm{\nabla R_n(\beta_0^{\star}, \M{0}, \phi^{\star})}_{\infty}$
denoting the smallest value of $\lambda$ for which $(\beta_0^{\star}, \M{0}, \phi^{\star})$ is a minimizer of 
the optimization problem \eqref{eq:estimator} according to the KKT conditions; here, $\beta_0^{\star}$ and $\phi^{\star}$ represent the values of the MLE of these parameters in a plain intercept model. 

For each value of $\lambda$, we run the proximal gradient method in $\S$\ref{subsec:optimization} with stepsize $s = .5 \cdot \nnorm{\nabla_{\beta_0, \beta}^2 R_n(\wh{\beta}_0(0), \wh{\beta}(0), \wh{\phi}(0)}$, where $(\wh{\beta}_0(0), \wh{\beta}(0), \wh{\phi}(0))$ denote the (unpenalized)
solution with $\lambda = 0$ obtained through the \textsf{betareg} package, which also serves
as initial solution. The scale parameter $\phi$ is updated every $M = 5$ iterations. 

In addition, we compute the solution for $\lambda = 2 \sqrt{\log(p)/n}$ that is roughly in line with the choice of the tuning parameter according to the theoretical analysis in $\S$\ref{sec:analysis}. Table \ref{tab:coef_estimates} lists the resulting coefficient estimates and their confidence
intervals where the matrix $\wh{\Omega}$ is computed with the implementation of \cite{Javanmard2014} available from \cite{sslasso}. As a reference, Table \ref{tab:coef_estimates} also lists the coefficient estimates of the AIC-optimal model obtained via exhaustive enumeration of 
all $2^p \approx 10^6$ variable subsets. 
\vskip1ex
\noindent {\em Discussion}. The AIC-optimal model produces a more parsimonious solution than $\ell_1$-penalization with the above choice of the tuning parameter (8 vs.~12 non-zero coefficients). The latter does not miss any of the non-zero variables of the AIC-optimal model and the sets of prominent variables with comparatively large coefficients (Poor physical health days, Health Care costs, Police officers, and \% Children in poverty) agree fully. As expected, the coefficients of the $\ell_1$-penalization approach undergo significant shrinkage towards zero, which likely explains the over-selection of extraneous variables to compensate for this shrinkage \cite{Zhao2006}. Observe that zero is contained in each confidence intervals for variables with a coefficient estimate of zero. At the same time, none of the confidence intervals of the variables
included in the AIC-optimal contain zero; all except for one of the intervals (the one for ``County size is large") include the coefficient estimate of the AIC-optimal model. Among the four extra variables included in the $\ell_1$-penalization model, the confidence intervals of two of these 
(``Psychiatrists" and ``\% Population over the age of 18") contain zero; for, the variables ``\% African American" and ``Median Household income" the zero is not included, which constitutes the only noticeable discrepancy from the AIC-optimal model. 

\begin{table}
\begin{center}
{\footnotesize  \begin{tabular}{|l|c|c|c|c|c|}
\hline
                  & Inc.~ineq.& HS grad.~rate & Size=medium &  Size=large & Poor Phys.  \\ 
                  \hline & & & & &\\[-1ex]
$\wh{\beta}_j$    &    0       & $-.0041$ & $-.0048$  & $-.0127$ & $.0683$ \\[.5ex]    
CI                &   {\scriptsize $[-.032, .032]$}        & {\scriptsize $[-.053, -.001]$}  & {\scriptsize $[-.065, -.015]$}    & {\scriptsize $[-.043, -.007]$} &  {\scriptsize $[.028, .173]$} \\[.5ex]
$\wh{\beta}_j^{\text{AIC}}$ & 0 & $-.0397$ & $-.0472$ & $-.0725$   & $.0764$ \\[.5ex]
\hline
\hline
& Poor Ment. & PCP  & HC cost  & \%Medicaid &  Psychiatrists   \\ 
                  \hline & & & & &\\[-1ex]
$\wh{\beta}_j$    & 0          & 0 &  $.0439$  & $-.0234$& $-.0057$ \\[.5ex]    
CI                & {\scriptsize $[-.062, .062]$} & {\scriptsize $[-.033, .033]$} & {\scriptsize $[.024, .095]$}   & {\scriptsize $[-.098, -.044]$} & {\scriptsize $[-.026, .055]$} \\[.5ex]
$\wh{\beta}_j^{\text{AIC}}$ & 0 & 0 & $.0751$ &  $-.0470$ & 0 \\[.5ex]
\hline
\hline
& Comm.~MHC & Violent crimes  & Police officers & Household Inc.  & \%African    \\ 
                  \hline & & & & &\\[-1ex]
$\wh{\beta}_j$    &  0         &  0 & $.0641$ & $-.0265$ & $.0187$ \\[.5ex]    
CI                &  {\scriptsize $[-.015, .015]$} & {\scriptsize $[-.046, .046]$}  & {\scriptsize $[.024, .121]$}  & {\scriptsize $[-.245, -.156]$} & {\scriptsize $[.018, .097]$} \\[.5ex]
$\wh{\beta}_j^{\text{AIC}}$ & 0  & 0  & $.0828$ &  0 & 0 \\[.5ex]
\hline
\hline
& \%Hispanic & \%Rural  & \%Over18 & \%ChildrenPov  & HPSA    \\ 
                  \hline & & & & &\\[-1ex]
$\wh{\beta}_j$    &  0         & 0 & $-.0117$ & $.0727$ & 0 \\[.5ex]    
CI                &  {\scriptsize $[-.034, .034]$}  & {\scriptsize $[-.042, .042]$}   & {\scriptsize $[-.040, .046]$}   & {\scriptsize $[.073, .214]$} & {\scriptsize $[-.027, .027]$} \\[.5ex]
$\wh{\beta}_j^{\text{AIC}}$ & 0 & 0  & 0  & $.0968$ & 0 \\[.5ex]
\hline
\end{tabular}}
\end{center}
\vspace*{-3ex}
\caption{Coefficient estimates $\{ \wh{\beta}_j \}$, their confidence intervals (note that by construction, these may not need to be centered at the parameter estimates), and the corresponding coefficient estimates $\{ \wh{\beta}_j^{\text{AIC}} \}$ in the AIC-optimal model. For space reasons, we here list the intercepts $\wh{\beta}_0 = -5.28$ and $\wh{\beta}_0^{\text{AIC}} = -5.29$ and scale parameters $\wh{\phi} = 375.3$ and $\wh{\phi}^{\text{AIC}} = 384.3$.}\label{tab:coef_estimates}
\end{table}
\vspace*{-1.5ex}
\section{Conclusion}\label{sec:conclusion}
\vspace*{-1.5ex}
In this paper, we have studied $\ell_1$-penalized Beta regression in a high-dimensional setup. We have provided a non-asymptotic bound on the $\ell_1$-estimation error of stationary points in 
a neighborhood of the true parameter, following the framework in \cite{Elsener2018} for 
$\ell_1$-penalized M-estimators with non-convex objective function. To enable inference for 
low-dimensional parameters such as individual regression coefficients, we 
adopt the debiasing approach in \cite{Zhang2014, vandegeer2014, Javanmard2014} that is
widely used for $\ell_1$-penalized estimators. We also present a scalable framework 
for optimization based on proximal gradient descent, a popular first-order algorithm. We apply
our approach to select variables in a Beta regression model for predicting the fraction of 
incarcerated individuals in US counties. This case study demonstrates the utility of our 
approach.

Looking ahead, there are various avenues for future research. First, it is worth examining 
the statistical properties of stationary points returned by the proposed proximal gradient
algorithm, thereby unifying computational and theoretical aspects. Second, in this paper we
have focused on the $\ell_1$-estimation error; studying the $\ell_2$-estimation error and 
prediction error (with respect to the loss induced by the negative log-likelihood) will complement 
the results in this paper. Third, it is worth considering concave penalties such as the 
SCAD \cite{Fan2001} and MCP \cite{Zhang2010} that reduce the bias affecting the estimation of the non-zero coefficients of the target. Fourth, in this paper we have focused on the homoscedastic
Beta regression model with a constant scale parameter. In heteroscedastic Beta regression, 
observation-specific scale parameters may depend on covariates. The paper \cite{Zhao2014} considers
sparse estimation in this setting when the mean and scale are modeled by two distinct linear predictors. It is worth investigating this scenario in a high-dimensional setup.

\clearpage  
\bibliographystyle{plain}
\bibliography{betareg}
\vskip8ex
\noindent {\Large {\bfseries \textsf{Appendix}}}

\subsection*{Proof of Lemma \ref{lem:SC}}
Let $v \in \R^p$ be arbitrary. Invoking Lemma \ref{lem:grad_hessian}, we have 
\begin{equation}\label{eq:SC:basic}
v^{\T} (\nabla^2 R(\beta) - \nabla^2 R(\beta^*)) v = \E[(X^{\T} v)^2 \{ \rho_X(X^{\T} \beta) - \rho_X(X^{\T} \beta^*) \}],
\end{equation}
where the map $\rho: \R \rightarrow \R$ is given by 
\begin{align*}
\rho_X(z) &= \{ \mu'(z) \}^2 \phi^{*2} \{ \Psi'(\mu(z) \phi^*) + \Psi'((1 - \mu(z)) \phi^*) \} \\
&\quad + \mu''(z) \phi^* \{ [\Psi(\mu(z) \phi^*) - \Psi(\mu(X^{\T} \beta^*) \phi^*)] -  \\
&\qquad [\Psi((1 - \mu(z)) \phi^*) - \Psi((1 - \mu(X^{\T} \beta^*)) \phi^*)] \}, \; z \in \R. 
\end{align*}
In the sequel, we will show that $\rho_X$ is Lipschitz on $[-h,h]$. For this purpose, we let $L_f$ denote the Lipschitz constant of some function $f$. Using the properties established in Lemmas \ref{lem:lipschitz} and \ref{lem:lipschitz_and_bounded}, we obtain that 
\begin{align*}
L_{\rho_X} &\leq 2 \phi^{*2} L_{\mu'}  \nnorm{\mu'}_{\infty} \, 2 L_{\Psi}  + \phi^{*2} \nnorm{(\mu')^2}_{\infty} \{ 2 L_{\Psi'} L_{\mu} \, \phi^* \} + 4 \phi^* L_{\mu''} \nnorm{\Psi}_{\infty} + 2 L_{\Psi'} \phi^* \nnorm{\mu''}_{\infty} \\
&\leq \frac{1}{4} \phi^{*2} C_{h,\phi}' + \frac{1}{32} \phi^{*3} C_{h, \phi}'' + \frac{3}{2} \phi^* C_{h,\phi} + \frac{1}{2} \phi^* C_{h,\phi}'' \invcoloneq \overline{L}_{h,\phi^*}. 
\end{align*}
Consequently, in light of \eqref{eq:SC:basic}, 
\begin{align*}
v^{\T} (\nabla^2 R(\beta) - \nabla^2 R(\beta^*)) v &\leq \overline{L}_{h,\phi^*} \E[(X^{\T} v)^2  |X^{\T} (\beta - \beta^*)| ] \\
&\leq \overline{L}_{h,\phi^*} \sqrt{\E[(X^{\T} v)^{4}]} \sqrt{\E[\{X^{\T} (\beta - \beta^*)\}^2]} \\
&\leq \overline{L}_{h,\phi^*} \nnorm{v}_2^2 \, 2^{3/2} C_X^{3} \nnorm{\beta - \beta^*}_2,
\end{align*}
using that $\E[(X^{\T} v /\nnorm{v})^4] \leq 4 C_X^4$ and $\E[\{ X^{\T} (\beta - \beta^*)/\nnorm{\beta - \beta^*}_2\}^2] \leq 2 C_X^2$ according to well-known bounds on the moments of $\psi_2$-random variables. The assertion of the lemma then follows from $v^{\T} R(\beta^*) v \geq \underline{\mu} \nnorm{v}_2^2$. \qed

\subsection*{Proof of Theorem \ref{highdim_consistency}}
The result of the Theorem is obtained verifying the conditions of Theorem \ref{theo:vdG}. We first
note that by assumption and according to Lemma \ref{lem:SC}, the risk $R$ is $\underline{\mu}/2$- strongly convex in $\mc{B} = \{\beta \in \R^p: \, \nnorm{\beta - \beta^*}_2 \leq \eta\}$. The rest of the proof is devoted towards demonstrating that 
\begin{equation}\label{eq:emp-process}
\sup_{\beta \in \mc{B}} \left| \{ \nabla R_n(\beta) - \nabla R(\beta) \}^{\T} (\beta - \beta^*)\right| \leq \lambda_{\eps} \nnorm{\beta - \beta^*}_1
\end{equation}
with high probability (to be specified), for a suitable choice of $\lambda_{\eps} > 0$, which will prompt the assertion of the theorem in virtue of Theorem \ref{theo:vdG}. Establishing 
\eqref{eq:emp-process} relies on empirical process techniques, following the basic template 
developed in \cite{Elsener2018}. That template is combined with an additional truncation step
to cope with the unboundedness of the terms $\{ \log(Y_i), \log(1- Y_i) \}_{i = 1}^n$ appearing
in $\nabla R_n$. Below, we list the main steps to obtain \eqref{eq:emp-process}.    
\begin{itemize}
\item[1.] Establish \eqref{eq:emp-process} for $\mc{B}_M = \{\beta \in \mc{B}:\, \nnorm{\beta - \beta^*}_1 \leq M   \}$ using the concentration inequality in Lemma \ref{lem:Bousquet} combined
with a truncation step.
\item[2.] Bound the individual terms appearing in Lemma \ref{lem:Bousquet} for the specific
          scenario under consideration here.
\item[3.] Extend the result from $\mc{B}_M$ to $\mc{B}$ via the peeling device \cite{Geer2000}. 
\end{itemize}

\noindent \underline{{\em Step 1}}.\\ 
Define the class of functions
\begin{align*}
&\bigg\{f_{\beta}(x,y) \coloneq \phi^* \cdot x^{\T} (\beta - \beta^*) \mu'(x^{\T}\beta) \times \\
&\times \left\{[\Psi(\mu(x^{\T} \beta) \phi^*) - \log(y)] - [\Psi( (1 -
\mu(x^{\T} \beta)) \phi^*)  - \log(1 - y)] \right\}, \;  (x,y) \in \R^{p} \times (0,1) \bigg \}_{\beta \in \R^p}. 
\end{align*}
Accordingly, we have 
\begin{align*}
&\nabla R_n(\beta)^{\T} (\beta - \beta^*) = \frac{1}{n} \sum_{i = 1}^n f_{\beta}(X_i, Y_i) \\
&\{\nabla R_n(\beta) - \nabla R(\beta) \}^{\T} (\beta - \beta^*) = \frac{1}{n} \sum_{i = 1}^n \{ f_{\beta}(X_i, Y_i) - \E[f_{\beta}(X_i, Y_i)] \}. 
\end{align*}
Let further 
\begin{equation*}
\M{Z}_M = \sup_{\beta \in \mc{B}_M} \left|\frac{1}{n}  \sum_{i = 1}^n \{ f_{\beta}(X_i, Y_i) - \E[f_{\beta}(X_i, Y_i)] \} \right|.
\end{equation*}
For $y \in (0,1)$, define the map $m(y) = \max\{ \log(y), \log(1-y) \}$, and let $\{ t_M \}$ and $u$ be arbitrary positive numbers. We then have 
{\small \begin{align}
  &\p \left(\exists M: \M{Z}_M > t_M \right) = \p \left(\exists M: \sup_{\beta \in \mc{B}_M} \left| \frac{1}{n} \su \{ f_{\beta}(X_i, Y_i) - \E[f_{\beta}(X_i, Y_i)]\}  \right| > t_M  \right) \notag \\
                                        &\leq  \p \Bigg(\exists M: \sup_{\beta \in \mc{B}_M} \Bigg \{ \left| \frac{1}{n} \su \{f_{\beta}(X_i, Y_i) \mathbf{1}_{\{m(Y_i) \leq u\}} - \E[f_{\beta}(X_i, Y_i)]\} \right|   \ \notag \\
                                        &\qquad \qquad \qquad \qquad +  \left| \frac{1}{n} \su  f_{\beta}(X_i, Y_i) \mathbf{1}_{\{m(Y_i) > u\}} \right|  > t_M  \Bigg\} \Bigg) \notag \\
                                        &\leq \p \Bigg(\exists M: \sup_{\beta \in \mc{B}_M}  \left| \frac{1}{n} \su \{ f_{\beta}(X_i, Y_i) \mathbf{1}_{\{m(Y_i) \leq u\}} - \E[f_{\beta}(X_i, Y_i)]\} \ \right|  \notag \\
                                        &\qquad \qquad +  \sup_{\beta \in \mc{B}_M} \left| \frac{1}{n} \su f_{\beta}(X_i, Y_i) \mathbf{1}_{\{m(Y_i) > u\}} \right|   > t_M  \Bigg) \notag \\
                                        &\leq \p \left(\exists M: \sup_{\beta \in \mc{B}_M}  \left| \frac{1}{n} \su \{ f_{\beta}(X_i, Y_i) \mathbf{1}_{\{m(Y_i) \leq u\}} - \E[f_{\beta}(X_i, Y_i)] \}\right|  > t_M  \right) +  \notag \\
  &\quad +\p \left(\exists M: \sup_{\beta \in \mc{B}_M}  \left| \frac{1}{n} \su f_{\beta}(X_i, Y_i) \mathbf{1}_{\{m(Y_i)  > u\}} \right|  > 0  \right) \label{eq:emp_truncation}
\end{align}}
Regarding the second term on the right hand side of the last inequality, we use the bound 
\begin{align}
  &\p \left(\exists M: \sup_{\beta \in \mc{B}_M}  \left| \frac{1}{n} \su f_{\beta}(X_i, Y_i) \mathbf{1}_{\{m(Y_i)  > u\}} \right|  > 0  \right) \notag \\
  &= 1 - \left(\forall M: \sup_{\beta \in \mc{B}_M}  \left| \frac{1}{n} \su f_{\beta}(X_i, Y_i) \mathbf{1}_{\{m(Y_i)  > u\}} \right|  = 0  \right) \notag \\
                                                                                                                                            &\leq  1 - \left(\bigcap_{i = 1}^n \mathbf{1}_{\{m(Y_i)  \leq u\}}   \right) \leq C_{\phi^*, h} \frac{1}{n^{\kappa - 1}}, \label{eq:term2_control}                                                                                                                 
\end{align}
for some constant $C_{\phi^*, h} > 0$ depending only on $\phi^*$ and $h$ (cf.~Assumption {\bfseries (A3)}) and $\kappa > 0$ to be chosen below,  where the last inequality in \eqref{eq:term2_control}  is obtained by invoking Lemma \ref{tail:beta} with the choice 
\begin{equation}\label{eq:tail_choice}
t = u = \frac{\kappa \log 2n}{\phi^* \min_{1 \leq i \leq n} \min \{ \mu_i(\beta^*), 1 - \mu_i(\beta^*) \}} \leq \frac{\kappa \log 2n}{c_h \phi^*} \invcoloneq \wt{C}_{h,\phi^*} \, \kappa \log 2n 
\end{equation}
and $c_h$ as defined in Assumption {\bfseries (A3)}, and then using 
the union bound over $1 \leq i \leq n$. 

Regarding the first term on the right hand side in \eqref{eq:emp_truncation}, we have 
{\small \begin{align}
&\p \left(\sup_{\beta \in \mc{B}_M}  \left| \frac{1}{n} \su \{f_{\beta}(X_i, Y_i) \mathbf{1}_{\{m(Y_i)  \leq u\}} - \E[f_{\beta}(X_i, Y_i)]\}   \right|  > t_M\right) \notag \\
&= \p \left(\sup_{\beta \in \mc{B}_M}  \left| \frac{1}{n} \su \{f_{\beta}(X_i, Y_i) \mathbf{1}_{\{m(Y_i)  \leq u\}} - \E[f_{\beta}(X_i, Y_i) \M{1}_{m(Y_i) \leq u}] - \E[f_{\beta}(X_i, Y_i) \M{1}_{m(Y_i) > u}]\}  \right|  > t_M  \right) \notag \\
&\leq \p \Bigg(\underbrace{\sup_{\beta \in \mc{B}_M}  \left| \frac{1}{n} \su \{f_{\beta}(X_i, Y_i) \M{1}_{m(Y_i) \leq u} - \E[f_{\beta}(X_i, Y_i) \M{1}_{m(Y_i) \leq u}] \} \right|}_{(I)} \notag \\
&\qquad + \underbrace{\sup_{\beta \in \mc{B}_M}  \left|\frac{1}{n} \su \E[f_{\beta}(X_i, Y_i) \M{1}_{m(Y_i) > u}] \right|}_{(II)} > t_M  \Bigg). \label{eq:term1_control} 
\end{align}}
Now observe that for (II), we have 
\begin{align}
\sup_{\beta \in \mc{B}_M} \left|\frac{1}{n} \su \E[f_{\beta}(X_i, Y_i) \M{1}_{m(Y_i) > u}] \right| &= \sup_{\beta \in \mc{B}_M} \left|\E[f_{\beta}(X, Y) \M{1}_{m(Y) > u}] \right| \notag \\
&\leq \sup_{\beta \in \mc{B}_M} \sqrt{\E[f_{\beta}^2(X, Y)]} \sqrt{\E[\M{1}_{m(Y) > u}]} \notag \\
&\leq n^{-\kappa/2} \sup_{\beta \in \mc{B}_M} \sqrt{\E[f_{\beta}^2(X, Y)]} \label{eq:Etruncbound_1}
\end{align}
by choosing $u$ as in \eqref{eq:tail_choice}. Furthermore, we have for any $\beta \in \mc{B}_M$
\begin{align}
\E[f_{\beta}^2(X, Y)] &\leq \frac{\phi^{*2}}{4} \E \left[\{ X^{\T} (\beta - \beta^*)\}^2 ( 4 \log(Y)^2 + 4\log(1 - Y)^2) \right] + \notag \\
&\quad + \frac{\phi^{*2}}{4} \bar{C}_{h,\phi^*} \E[\{ X^{\T} (\beta - \beta^*)\}^2] \label{eq:Etruncbound_2}
\end{align}
where $\bar{C}_{h,\phi^*} = 4 \nnorm{\Psi}_{\infty} = 4 C_{h,\phi^*}$ per Lemma \ref{lem:lipschitz_and_bounded}. 
Furthermore, we have 
\begin{align}
&\E \left[\{ X^{\T} (\beta - \beta^*)\}^2 ( 4 \log(Y)^2 + 4\log(1 - Y)^2) \right] \notag \\
&= \E \left[\{ X^{\T} (\beta - \beta^*)\}^2 \E\left[( 4 \log(Y)^2 + 4\log(1 - Y)^2) | X \right]\right] \notag \\ 
&\leq \check{C}_{\phi^*, h} \E \left[\{ X^{\T} (\beta - \beta^*)\}^2 \right], \label{eq:Etruncbound_3}
\end{align}
for some constant $\check{C}_{\phi^*, h} > 0$ depending only on $\phi^*$ and $h$, using the moment bounds for log-beta random variables in Lemma \ref{tail:beta}. We then use that 
for all $\beta \in \mc{B}_M$, we have 
\begin{equation}\label{eq:Etruncbound_4}
\E[ X^{\T}(\beta - \beta^*) \}^2] \leq 2C_X \cdot M^2,
\end{equation}
where $C_X = \nnorm{X}_{\psi_2}$. Combining \eqref{eq:Etruncbound_1} through \eqref{eq:Etruncbound_4}, we obtain that 
\begin{equation}\label{eq:Etruncbound_final}
\sup_{\beta \in \mc{B}_M} \left|\frac{1}{n} \su \E[f_{\beta}(X_i, Y_i) \M{1}_{m(Y_i) > u}] \right| \leq n^{-\kappa/2} \, \overline{C}_{\phi^*, h} \cdot C_X \cdot M,
\end{equation}
for $\kappa > 0$ to be chosen below. 
\vskip2ex
\noindent \underline{{\em Step 2}}.\\  
We now turn our attention to term (I) in \eqref{eq:term1_control}, i.e., the supremum of an empirical process, to be targeted with Lemma \ref{lem:Bousquet}. We start by controlling the expectation of the supremum of the empirical process using Rademacher symmetrization \cite[e.g.,][]{Wainwright2019} and the contraction principle in the form of Lemma \ref{lem:contraction}. Let 
$\{ \eps_i \}_{i = 1}^n$ be an i.i.d.~sequence of Rademacher random variables independent 
of $\{ (X_i, Y_i) \}_{i = 1}^n$. We then obtain
\begin{align}\label{eq:bound_expectation_of_emp}
&\E \left[\sup_{\beta \in \mc{B}_M}  \left| \frac{1}{n} \su \{f_{\beta}(X_i, Y_i) \M{1}_{m(Y_i) \leq u} - \E[f_{\beta}(X_i, Y_i) \M{1}_{m(Y_i) \leq u}] \right| \right] \notag \\
&\leq 2 \E_{\{ (X_i, Y_i, \eps_i) \}} \left[\sup_{\beta \in \mc{B}_M}  \left| \frac{1}{n} \su \eps_i f_{\beta}(X_i, Y_i) \M{1}_{m(Y_i) \leq u} \right| \right]
\end{align}
To apply the contraction principle in Lemma \ref{lem:contraction}, we condition on the 
$\{ (X_i, Y_i) \}_{i = 1}^n$ and then apply the Lemma with 
(i) $z_i = X_i$, $1 \leq i \leq n$, (ii) the class of functions $\mc{G} = \{z \mapsto g_{\beta}(z) = z^{\T} \beta, \; \beta \in \mc{B}_M \}$, (iii) the functions $\gamma_i: [-h,h] \rightarrow \R$ defined by
\begin{align*}
t \mapsto \gamma_i(t) &= \big( \phi^* t \mu'(t) \left \{ \left[ \Psi(\mu(t) \phi^*) - \log(Y_i)  \right]  - \left[ \Psi((1 - \mu(t)) \phi^*) - \log(1 - Y_i)  \right] \right\} \\
& - \phi^* X_i^{\T} \beta^* \mu'(t) \left \{ \left[ \Psi(\mu(t) \phi^*) - \log(Y_i)  \right]  - \left[ \Psi((1 - \mu(t)) \phi^*) - \log(1 - Y_i)  \right] \right\} \big) \mathbf{1}_{m(Y_i) \leq u}, 
\end{align*}
$1 \leq i \leq n$, and (iv) $g^* = g_{\beta^*}$ (observe that $\gamma_i(g_{\beta^*}(z_i)) = 0$,
$1 \leq i \leq n$). Accordingly, 
\begin{align}\label{eq:bound_expectation_of_emp_cont}
&\E_{\{ (X_i, Y_i, \eps_i) \}} \left[\sup_{\beta \in \mc{B}_M}  \left| \frac{1}{n} \su \eps_i f_{\beta}(X_i, Y_i) \M{1}_{m(Y_i) \leq u} \right| \right] \notag \\
&= \E_{\{ (X_i, Y_i) \}} \left[ \E_{\{ \eps_i \}_{i = 1}^n}\left[\sup_{\beta \in \mc{B}_M}  \left| \frac{1}{n} \su \eps_i f_{\beta}(X_i, Y_i) \M{1}_{m(Y_i) \leq u} \right| \; \Bigg| \{(X_i, Y_i) \}_{i = 1}^n \right] \right] \notag \\
&= \E_{\{ (X_i, Y_i) \}} \left[ \E_{\{ \eps_i \}_{i = 1}^n}\left[\sup_{\beta \in \mc{B}_M}  \left| \frac{1}{n} \su \eps_i \gamma_i(g_{\beta}(X_i)) \right| \; \Bigg| \{(X_i, Y_i) \}_{i = 1}^n \right] \right] \notag \\
&= \E_{\{ (X_i, Y_i) \}} \left[ \E_{\{ \eps_i \}_{i = 1}^n}\left[\sup_{\beta \in \mc{B}_M}  \left| \frac{1}{n} \su \eps_i \{ \gamma_i(g_{\beta}(X_i)) - \gamma_i(g^*(X_i))\}\right| \; \Bigg| \{(X_i, Y_i) \}_{i = 1}^n \right] \right] \\
&\leq  2 \E_{\{ (X_i, Y_i) \}} \left[\max_{1 \leq i \leq n} L_{\gamma_i} \E_{\{ \eps_i \}_{i = 1}^n}\left[\sup_{\beta \in \mc{B}_M}  \left| \frac{1}{n} \su \eps_i \{ g_{\beta}(X_i) - g^*(X_i)\}\right| \; \Bigg| \{(X_i, Y_i) \}_{i = 1}^n \right] \right], \notag
\end{align}
where the inequality is obtained from Lemma \ref{lem:contraction}, with $L_{\gamma_i}$ denoting the Lipschitz constant of $\gamma_i$, $1 \leq i \leq n$. In the sequel, we will bound these Lipschitz constants. First note that if $m(Y_i) > u$, then $\gamma_i \equiv 0$, $1 \leq i \leq n$ and there is nothing to show. In the opposite case, we differentiate $t \mapsto \gamma_i(t)$, and bound the resulting derivative $t \mapsto \gamma_i'(t)$ on $[-h,h]$. For $1 \leq i \leq n$, letting $c_i = X_{i}^{\T} \beta^*$, we have
\begin{align}\label{eq:lipschitz_gamma}
  \gamma_i'(t) &= \phi^* \underbrace{\mu'(t) \Psi(\mu(t) \phi^*)}_{\leq \frac{1}{4} C_{h,\phi^*}}  + \phi^* \underbrace{t \mu''(t) \Psi(\mu(t) \phi^*)}_{\leq \frac{h}{4} C_{h,\phi^*}} + \phi^{*2} \underbrace{t \Psi'(\phi^* \mu(t)) \{ \mu'(t) \}^2}_{\leq \frac{h}{16} C_{h,\phi^*}'}  \notag \\
            &-\phi^* \underbrace{\log(Y_i) \mu'(t)}_{\leq u/4} - 
            \phi^* \underbrace{t \log(Y_i) \mu''(t)}_{\leq \frac{h \cdot u}{4}} \notag \\
            &-\phi^* \underbrace{c_i \Psi(\mu(t) \phi^*) \mu''(t)}_{\leq \frac{h}{4} C_{h,\phi^*}} - \phi^{*2} \underbrace{c_i \Psi'(\mu(t) \phi^*) \{ \mu'(t) \}^2}_{\leq \frac{1}{16} C_{h,\phi^*}'}  \notag \\
            &+\phi^* \underbrace{c_i \mu''(t) \log(Y_i)}_{\leq \frac{h \cdot u}{4}}  \notag \\
            &-\phi^* \underbrace{\mu'(t) \Psi((1-\mu(t)) \phi^*)}_{\leq C_{h,\phi^*}/4} - \phi^* \underbrace{t \mu''(t) \Psi((1 - \mu(t) \phi^*))}_{\leq \frac{h}{4} C_{h,\phi^*}} + \phi^{*2} \underbrace{t \Psi'((1- \mu(t)) \phi^*) \{ \mu'(t) \}^2}_{\leq \frac{h}{16} C_{h,\phi}'} \notag \\
            &+\phi^* \underbrace{\mu'(t) \log(1 - Y_i)}_{\leq \frac{u}{4}} + \phi^* \underbrace{ t \mu''(t) \log(1 - Y_i)}_{\leq \frac{h \cdot u}{4}} \notag \\
            &+ \phi^* \underbrace{c_i \mu''(t) \Psi(\phi^* (1 - \mu(t)))}_{\leq \frac{h}{4} C_{h,\phi^*}} - \phi^{*2} \underbrace{c_i \Psi'((1 - \mu(t)) \phi^*)  \{ \mu'(t) \}^2 }_{\leq \frac{h}{16} C_{h,\phi^*}'} \notag \\
            &- \phi^* \underbrace{c_i \mu''(t) \log(1 - Y_i)}_{\leq \frac{h \cdot u}{4}},  \quad t \in [-h,h]. \notag \\
            &\leq  u (h+ 1/2) + C_{h,\phi^*} (h + 1/2) + h C_{h,\phi^*}' \phi^*/4 \leq K_{h,\phi^*} \kappa \log 2n. 
\end{align}
where in the above display, all inequalities (which follow from Assumption {\bfseries (A3)} and Lemma \ref{lem:lipschitz_and_bounded}) apply to the terms in absolute value. In the last inequality, we have inserted the choice for $u$ in \eqref{eq:tail_choice} and $K_{h, \phi^*} > 0$ is a constant depending only on $h, \phi^*$. 

Combining \eqref{eq:bound_expectation_of_emp}, \eqref{eq:bound_expectation_of_emp_cont}, 
and \eqref{eq:lipschitz_gamma}, we obtain 
\begin{align}\label{eq:Ebound_final}
&\E \left[\sup_{\beta \in \mc{B}_M}  \left| \frac{1}{n} \su \{f_{\beta}(X_i, Y_i) \M{1}_{m(Y_i) \leq u} - \E[f_{\beta}(X_i, Y_i) \M{1}_{m(Y_i) \leq u}] \right| \right] \notag \\
&\leq 4 K_{h, \phi^*} \kappa \log(2n)  \E_{\{ (X_i, Y_i) \}} \left[ \E_{\{ \eps_i \}}\left[\sup_{\beta \in \mc{B}_M} \left| \frac{1}{n} \su \eps_i X_i^{\T} (\beta - \beta^*)\right|  \; \Bigg| \{(X_i, Y_i) \}_{i = 1}^n \right] \right] \notag \\
&\leq 4 K_{h, \phi^*} \kappa \log(2n) \times M \cdot \E\left[\norm{ \frac{1}{n} \su \eps_i X_i}_{\infty} \right] \notag \\
&\leq 4 K_{h, \phi^*} \kappa \log(2n) \times M \cdot \left\{ \nnorm{(\eps_i)}_{\psi_2} \nnorm{(X_{ij})}_{\psi_2} \frac{\log(p+1)}{n} + \sqrt{\frac{4 \nnorm{(\eps_i)}_{\psi_2}^2 \nnorm{(X_{ij})}_{\psi_2}^2 \log(p + 1)}{n}} \right \} \notag \\
&\invcoloneq M \wt{\lambda}_1, 
\end{align}
where the last inequality can be taken directly from \cite[][cf.~Appendix D.5]{Elsener2018}.

To apply Bousquet's concentration inequality (Lemma \ref{lem:Bousquet}) to control the term (I) in \eqref{eq:term1_control}, it remains to determine $R > 0$ such that $\E[\{ f_{
\beta}(X_1, Y_1) \M{1}_{m(Y_1) \leq u} - \E[f_{
\beta}(X_1, Y_1) \M{1}_{m(Y_1) \leq u}] \}^2] \leq R^2$ and $K > 0$ such that 
$|\{ f_{
\beta}(X_1, Y_1) \M{1}_{m(Y_1) \leq u} - \E[f_{
\beta}(X_1, Y_1) \M{1}_{m(Y_1) \leq u}] \}| \leq K$ almost surely for all $\beta \in \mc{B}_M$. Regarding the former, we have 
\begin{align}\label{eq:bound_Rsq}
&\E[\{ f_{
\beta}(X_1, Y_1) \M{1}_{m(Y_1) \leq u} - \E[f_{
\beta}(X_1, Y_1) \M{1}_{m(Y_1) \leq u}] \}^2] \notag \\
&\leq \E[f^2_{
\beta}(X_1, Y_1) \M{1}_{m(Y_1) \leq u}] \notag \\
&\leq \E\bigg[\phi^{*2} \{ X_1^{\T} (\beta - \beta^*) \}^2 \{ \mu'(X_1^{\T} \beta) \}^2
    \Big\{[\Psi(\mu(X_1^{\T} \beta) \phi^*) - \log(Y_1)] \notag \\
    &\qquad \qquad \qquad \qquad \qquad \qquad \qquad \qquad - [\Psi( (1 -
  \mu(X_1^{\T} \beta)) \phi^*)  - \log(1 - Y_1)] \Big \}^2 \M{1}_{m(Y_1) \leq u} \bigg] \notag \\
&\leq  \E\bigg[ \phi^{*2} \{ X_1^{\T} (\beta - \beta^*) \}^2 \{ \mu'(X_1^{\T} \beta) \}^2 \times \notag \\
    &\qquad \quad 4\Big\{[\Psi^2(\mu(X_1^{\T} \beta) \phi^*) + \log^2(Y_1)] + [\Psi( (1 -
      \mu(X_1^{\T} \beta)) \phi^*)^2  + \log^2(1 - Y_1)] \Big \} \M{1}_{m(Y_1) \leq u} \bigg] \notag \\
&\leq \frac{4}{16} \phi^{*2} [ 2 \kappa^2 \log^2(2n) + 2 (C_{h,\phi^*}')^2] \E[\{ X_1^{\T} (\beta - \beta^*) \}^2]  \notag \\
&= \frac{4}{16} \phi^{*2} [ 2 \kappa^2 \log^2(2n) + 2 (C_{h,\phi^*}')^2] \E\left [ \left\{ X_1^{\T} \frac{(\beta - \beta^*)}{\nnorm{\beta - \beta^*}_2} \right \}^2 \right] \nnorm{\beta - \beta^*}_2^2 \notag \\
&\leq M^2 \phi^{*2} [\kappa^2 \log^2(2n) + (C_{h,\phi^*}')^2] \underbrace{\nnorm{X}_{\psi_2}^2}_{\leq C_X^2} \invcoloneq R^2, 
\end{align}
where we have inserted the choice for $u$ as above, and we have invoked Lemma \ref{lem:lipschitz_and_bounded} again. Similarly, we find that almost surely
\begin{align}\label{eq:bound_K}
|\{ f_{
\beta}(X_1, Y_1) \M{1}_{m(Y_1) \leq u} - \E[f_{
\beta}(X_1, Y_1) \M{1}_{m(Y_1) \leq u}] \}| &\leq 2 h \phi^* \{ C_{h, \phi^*}' + u \} \\
&\leq  2 h \phi^* \{ C_{h, \phi^*}' + \kappa \log(2n) \}  \notag         
\end{align}
Combining \eqref{eq:Ebound_final}, \eqref{eq:bound_Rsq}, \eqref{eq:bound_K},  \eqref{eq:Etruncbound_final} and \eqref{eq:term1_control}, an application of Lemma \ref{lem:Bousquet} with the choice
\begin{equation*}
t_M = M  \left\{2 \wt{\lambda}_1 + \phi^* [\kappa \log(2n) + C_{h,\phi^*}' C_X] \sqrt{\frac{t}{n}}   + n^{-\kappa/2} \overline{C}_{\phi^*, h} C_X \right \} + \frac{8 h \phi^* (C_{h,\phi^*}' + \kappa \log(2n))}{3} \frac{t}{n}
\end{equation*}
for any $t > 0$ yields
\begin{equation}\label{eq:term1_app_Bousq}
\p \Bigg(\sup_{\beta \in \mc{B}_M}  \underbrace{\left| \frac{1}{n} \su \{f_{\beta}(X_i, Y_i) \mathbf{1}_{\{m(Y_i)  \leq u\}} - \E[f_{\beta}(X_i, Y_i)]\}   \right|}_{\invcoloneq \M{Z}(\beta)}  > t_M \Bigg) \leq \exp(-t). 
\end{equation}
Defining 
\begin{equation*}
\wt{\lambda}_2(t) = \left\{2 \wt{\lambda}_1 + \phi^* [\kappa \log(2n) + C_{h,\phi^*}' C_X] \sqrt{\frac{t}{n}}   + n^{-\kappa/2} \overline{C}_{\phi^*, h} C_X \right \}
\end{equation*}
we can express \eqref{eq:term1_app_Bousq} as 
\begin{equation}\label{eq:term1_app_Bousq_sh}
\p \left(\sup_{\beta \in \mc{B}_M} \M{Z}(\beta) \geq M \wt{\lambda}_2(t) + \frac{8 h \phi^* (C_{h,\phi^*}' + \kappa  \log(2n))}{3} \frac{t}{n} \right) \leq \exp(-t). 
\end{equation}

\noindent \underline{{\em Step 3}}.\\ 
Note that the tail bound \eqref{eq:term1_app_Bousq_sh} concerns a fixed value of $M$. The goal of this step is to obtain a tail bound for any ``valid" $M$ to extend the analysis from the set 
$\mc{B}_M$ to the set of interest $\mc{B} = \{\beta \in \R^p: \; \nnorm{\beta - \beta^*}_2 \leq \eta\}$. Observe that $\sup_{\beta \in \mc{B}} \nnorm{\beta - \beta^*}_1 \leq \eta \sqrt{p}$, thus it suffices to consider all $M \in [0, \eta \sqrt{p}]$. Consider the ``slices" 
\begin{align*}
  S_0 = \left\{\beta \in \mc{B}:\, \nnorm{\beta - \beta^*}_1 \leq \frac{1}{n}  \right \}, \quad
  S_j &= \left\{\beta \in \mc{B}:\, \frac{2^{j-1}}{n} < \nnorm{\beta - \beta^*}_1 \leq \frac{2^j}{n}  \right \}, \\
&j = 1,\ldots, \lceil \log_2 (\eta n \sqrt{p}) \rceil, 
\end{align*}  
and observe that $\mc{B} \subseteq \bigcup_{j = 0}^{\lceil \log_2 (\eta n \sqrt{p}) \rceil} S_j$. We start with the slice $S_0$. In view of \eqref{eq:term1_app_Bousq_sh}, we have 
\begin{equation*}
\p\left(\exists \beta \in S_0: \; \M{Z}(\beta) \geq \frac{\wt{\lambda}_2(t)}{n} +  \frac{8 h \phi^* (C_{h,\phi^*}' + \kappa  \log(2n))}{3} \frac{t}{n} \right) \leq \exp(-t),
\end{equation*}  
Similarly, for each of the other slices, we obtain that
\begin{equation*}
\p\left(\exists \beta \in S_j: \; \M{Z}(\beta) \geq \frac{\wt{\lambda}_2(t) 2^j}{n} +  \frac{8 h \phi^* (C_{h,\phi^*}' + \kappa  \log(2n))}{3} \frac{t}{n}  \right) \leq \exp(-t),
\end{equation*}
Using the union bound and the choice $t = \log p$, we obtain that
\begin{align*}
\p\left(\exists \beta \in \mc{B}: \; \M{Z}(\beta) \geq  \nnorm{\beta - \beta^*}_1 \wt{\lambda}_2(t)\devs{t = \log p}  +  \frac{8 h \phi^* (C_{h,\phi^*}' + \kappa  \log(2n))}{3} \frac{\log p}{n}  \right) \leq \frac{\log_2(\eta n \sqrt{p})}{p},
\end{align*}
which can be spelled out as
\begin{align}\label{eq:peeling_end}
  \p\Bigg(\exists \beta \in \mc{B}: \; \M{Z}(\beta) \geq  \nnorm{\beta - \beta^*}_1  &\left[2 \wt{\lambda}_1 + n^{-\kappa/2} \overline{C}_{\phi^*, h} C_X + \phi^* [\kappa \log(2n) + C_{h,\phi^*}' C_X] \sqrt{\frac{\log p}{n}} \right] \notag \\
  &+ \frac{8 h \phi^* (C_{h,\phi^*}' + \kappa  \log(2n))}{3} \frac{\log p}{n}  \Bigg) \leq \frac{\log_2(\eta n \sqrt{p})}{p}. 
\end{align}

\noindent \underline{{\em Putting together the pieces}}.\\ 
Combining \eqref{eq:emp_truncation}, \eqref{eq:term2_control}, \eqref{eq:term1_control}, \eqref{eq:term1_app_Bousq_sh} and the argument of the previous section leading to \eqref{eq:peeling_end}, we obtain that with the choice $\kappa = 2$, the event 
\begin{align}\label{eq:lambda_eps}
\Big \{ &\left| \{ \nabla R_n(\beta) - \nabla R(\beta) \}^{\T} (\beta - \beta^*)\right| \notag \\
&\leq \nnorm{\beta - \beta^*}_1 \bigg( 2 C_0 C_X \frac{\log(p+1)}{n}  + 4 C_X \sqrt{\frac{\log (p+1)}{n}} + \frac{\overline{C}_{\phi^*, h} C_X}{n}  + \notag \\
&\; + \phi^* [2 \log(2n) + C_{h,\phi^*}' C_X] \sqrt{\frac{\log p}{n}} + \frac{8 h \phi^* (C_{h,\phi^*}' + 2 \log(2n))}{3} \frac{\log p}{n} \bigg) \quad\; \forall \beta \in \mc{B} \Big\}
\end{align}
occurs with probability at least $1 - C_{\phi^*, h}/n - \log_2(\eta n \sqrt{p})/p$. Using Lemma \ref{lem:SC} as indicated at the beginning of this proof, we can thus invoke Theorem \ref{theo:vdG} 
with $\mu = \underline{\mu}/2$ and $\lambda_{\eps}$ equal to the term following 
$\nnorm{\beta - \beta^*}_1$ in the right hand side of the inequality in \eqref{eq:lambda_eps} to conclude the proof. \qed

\renewcommand{\thelemma}{A.\arabic{lemma}}
\renewcommand{\thetheo}{A.\arabic{theo}}
\setcounter{theo}{0}
\setcounter{lemma}{0}
\subsection*{Auxiliary Results}
\begin{theo}\label{theo:vdG}(A simplified version of Theorem 2.1 in \cite{Elsener2018})\\
Let $\wh{\beta}$ be a stationary point of \eqref{eq:Rn_beta} in an $\eta$-neighborhood $\mc{C}$ of $\beta^*$, and suppose that the risk $R$ is $\mu$-strongly convex over $\mc{C}$. Let $\lambda_{\eps} > 0$ and $\lambda_* \geq 0$ such that for all  $\beta \in \mc{C}$ it holds that
\begin{equation*}
\left| \{ \nabla R_n(\beta) - \nabla R(\beta) \}^{\T} (\beta - \beta^*)\right| \leq \lambda_{\eps} \nnorm{\beta - \beta^*}_1 + \frac{\mu}{2} \nnorm{\beta - \beta^*}_2^2  + \lambda_*. 
\end{equation*}  
Let $\lambda > \lambda_{\eps}$, and define $\lambda_{-} = \lambda - \lambda_{\eps}$, $\lambda_+ = \lambda + \lambda_{\eps} + \frac{1}{2} \lambda_{-}$. Then we have
\begin{align*}
\nnorm{\wh{\beta} - \beta^*}_1 \leq 2 \left(\frac{1}{2\mu} \frac{\lambda_+^2}{\lambda_-} s  + \frac{\lambda_{*}}{\lambda_-} \right)
\end{align*} 

\begin{lemma}\label{lem:grad_hessian}(Gradient and Hessian of $R_n$ and $R$).\\
We have 
\begin{align*}
\nabla R_n(\beta) =&  \phi^* \frac{1}{n} \su X_i \mu_i'(\beta)
\left\{[\Psi(\mu_i(\beta) \phi^*) - \log(Y_i)] - [\Psi( (1 -
\mu_i(\beta)) \phi^*)  - \log(1 - Y_i)] \right \}   \\
\nabla R(\beta) =& \phi^*  \E\Big[X \mu'(X^{\T} \beta) \Big\{[\Psi(\mu(X^{\T} \beta) \phi^*) - \Psi(\mu(X^{\T} \beta^*) \phi^*)] - 
\\ 
&\qquad [\Psi( (1 -
\mu(X^{\T} \beta)) \phi^*)  - \Psi( (1- \mu(X^{\T} \beta^*)) \phi^*)] \Big \} \Big] \\
\nabla^2 R(\beta) =& \E\Big[X X^{\T}  \Big\{   \{ \mu'(X^{\T} \beta) \}^2
\phi^{*2} \{ \Psi'(\mu(X^{\T} \beta) \phi^*) + \Psi'((1 - \mu(X^{\T} \beta)) \phi^*) \}  + \\
&\qquad \mu''(X^{\T} \beta) \phi^* \{ [\Psi(\mu(X^{\T} \beta) \phi^*) -
  \Psi(\mu(X^{\T} \beta^*) \phi^*)] \\
&\qquad \qquad \qquad - [\Psi((1 - \mu(X^{\T} \beta)) \phi^*) -
  \Psi( (1- \mu(X^{\T} \beta^*)) \phi^*)] \} \Big\} \Big]
\end{align*}
\end{lemma}

\begin{bew} The gradient $\nabla R_n(\beta)$ is obtained by direct calculation, noting 
that the digamma function $\Psi$ equals the derivative of $\log \Gamma$. To obtain
$\nabla R(\beta) = \E[\nabla R_n(\beta)]$, we use that $\E[\log(Y_i)|X_i]= \Psi(\mu_i(\beta^*) \phi^*)$ and $\E[\log(1 - Y_i)|X_i] = \Psi((1 - \mu_i(\beta^*)) \phi^*)$, $1 \leq i \leq n$, in light of properties of the log-Beta distribution \cite{Gupta2004}. The expression for $\nabla^2 R(\beta)$ can be obtained from $\nabla R(\beta)$ by differentiating under the integral sign.  
\end{bew}

\begin{lemma}\label{tail:beta}(Tail bound and second moments for log-Beta random variables)\\ 
Let $Y \sim \textsf{{\em Beta}}(\mu \phi, (1 - \mu) \phi)$ for $\mu \in (c_h,1 - c_h)$ with $c_h = 1/(1 + \exp(h))$ for $h > 0$ and $\phi > 0$. Consider $U = -\log(Y)$ and $W = -\log(1 - Y)$. Then, for any $t \geq 1$
\begin{alignat}{2}
&\p(U > t) \leq C_{\phi, h} \exp(-t \mu \phi), \qquad &&\p(W > t) \leq C'_{\phi, h} \exp(-t (1 - \mu) \phi) \notag, \\
&\E[U^2] \leq C_{\phi, h}'', \qquad &&\E[W^2] \leq C_{\phi, h}''' \notag
\end{alignat}
for constants $C_{\phi, h}, C_{\phi, h}', C_{\phi, h}'', C_{\phi, h}''' > 0 $ depending only on $\phi$ and $h$.  
\end{lemma}
\begin{bew} The probability density function of $U$ is given by 
\begin{equation*}
f_U(u) = \frac{\Gamma(\phi)}{\Gamma(\mu \phi) \, \Gamma((1 - \mu) \phi)} [\exp(-u)^{\mu \phi - 1} (1 - \exp(-u))^{\phi (1 - \mu) - 1}] \exp(-u), \; u > 0.
\end{equation*}
Consequently, $f_U(u) \leq \wt{C}_{\phi, \mu} \exp(-u \mu \phi)$ with $\wt{C}_{\phi, \mu} = \frac{\Gamma(\phi)}{\Gamma(\mu \phi) \, \Gamma((1 - \mu) \phi)} \max\{(1-e^{-1})^{\phi(1-\mu) - 1}, 1\} \leq \wt{C}_{\phi, h}$ for any $u \geq 1$. Therefore, for $t \geq 1$ 
\begin{equation*}
\p(U \geq t) = \int_{t}^{\infty} f_U(u) \, du \leq \wt{C}_{\phi, h} \frac{1}{\mu \phi} \exp(-\mu \phi t) \leq C_{\phi, h} \exp(-\mu \phi t).
\end{equation*}
It follows that 
\begin{align*}
\E[U^2] = \int_{0}^{\infty} \p(U^2 \geq u) \, du &\leq 1 + \int_{1}^{\infty} \p(U \geq \sqrt{u}) \, du \\
&\leq 1 +   \int_{1}^{\infty} C_{\phi, h} \exp(-\mu \phi \sqrt{u}) \; du \\
          &\leq 1 +   \int_{1}^{\infty} C_{\phi, h} \, 2z \, \exp(-\mu \phi z) \; dz \\
          &\leq 1 +   C_{\phi, h}'  \frac{1}{\mu \phi} \int_{0}^{\infty} (\mu \phi) z \exp(-\mu \phi z) \; dz \\
  &\leq 1 +   C_{\phi, h}'  \frac{1}{(\mu \phi)^2} \leq C_{\phi, h}'',  
\end{align*}
where we have applied the change of variables $z = \sqrt{u}$ and used the fact that the expectation of 
an (Exponential) random variable with PDF $z \mapsto \exp(-\lambda z)$, $z,\lambda > 0$, is given by $1/\lambda$. Moreover, we have set $C_{\phi, h}' = 2 C_{\phi, h}$.

\noindent The corresponding inequalities for $W$ can be established similarly, noting that $1 - Y \sim \textsf{Beta}((1 - \mu) \phi, \mu \phi)$.
\end{bew}

\begin{lemma}(General Lipschitz properties).\label{lem:lipschitz}\\
Let $f: \R \rightarrow \R$ and $g: \R \rightarrow \R$ be bounded Lipschitz functions with Lipschitz constants 
$L_f$ and $L_g$, respectively. Then $f \circ g$ and $f \cdot g$ are also Lipschitz functions with Lipschitz constants $L_f L_g$ and $\nnorm{f}_{\infty} L_g + L_f \nnorm{g}_{\infty}$, respectively. 
\end{lemma}
\begin{bew} Let $x, x' \in \R$.  
We have $|f(g(x)) - f(g(x'))| \leq L_f |g(x) - g(x')| \leq L_f L_g |x - x'|$. Regarding the second property, observe that  
\begin{align*}
|f(x) g(x) - f(x') g(x')| &= |(f(x) - f(x')) g(x)  + f(x') (g(x) -
g(x'))|  \\
&\leq L_f \nnorm{g}_{\infty} |x - x'| + L_g \nnorm{f}_{\infty} |x - x'|. 
\end{align*}  
\end{bew}

\begin{lemma}(Lipschitz and boundedness properties).\label{lem:lipschitz_and_bounded}\\
With some abuse of notation, let $\mu(\beta)  = \mu(x^{\T} \beta) = \exp(x^{\T} \beta)/\{ 1+ \exp(x^{\T} \beta) \}$ and \linebreak $\mu'(\beta) = \frac{d \mu(z)}{d z} \dev{z}{x^{\T} \beta}$. Let further $\Psi(z) = \frac{d}{dz} \log \Gamma(z), \, z > 0$, be the Digamma function and let $\Psi'$ and $\Psi''$ denote its first and second derivative, respectively, restricted to the interval 
$[c_h \phi, \phi)$ with $c_h = (1 + \exp(h))^{-1}$ for $h > 0$.  We have 
\begin{align*}
&(i)\;0 \leq \mu'(\beta) \leq \frac{1}{4} \;\, \text{and} \;\, |\mu(\beta) - \mu(\beta')| \leq \frac{1}{4} |x^{\T} (\beta - \beta')|, \quad |\mu''(\beta)| \leq \frac{1}{4}, \quad |\mu'''(\beta)| \leq \frac{3}{8}.\\
&(ii) \nnorm{\Psi}_{\infty} \leq C_{h,\phi}, \quad L_{\Psi} = \Psi'(c_h \phi) = C_{h, \phi}' < \infty, \quad L_{\Psi'} = |\Psi''(c_h \phi)| = C_{h,\phi}'' < \infty,
\end{align*}
where $L_{\Psi}$ and $L_{\Psi'}$ denotes the Lipschitz constants of $\Psi$ and $\Psi'$, respectively, when restricted to $[c_h \phi, \phi)$. 
\end{lemma}
\begin{bew} One computes that $\mu'(\beta) = \mu(\beta) \cdot (1 - \mu(\beta))$, which yields the first part in (i). Furthermore, we have $\mu''(\beta) = \mu'(\beta)(1 - 2\mu(\beta))$ whose range 
is $[-1/4, 1/4]$. Similarly, $\mu'''(\beta) = \mu'(\beta) (1 - 2 \mu(\beta))^2 -2 \mu'(\beta)^2$ so that $\mu'''(\beta) \leq \frac{1}{4} + \frac{2}{16} = \frac{3}{8}$. Regarding part (ii), observe that $\Psi$ is monotonically increasing with $\lim_{z \rightarrow \infty} \Psi(z) = \infty$, and an asymptote at zero such that $\lim_{z \rightarrow 0} \Psi(z) = -\infty$. Therefore, $\nnorm{\Psi}_{\infty} = \max\{|\Psi(c_h \phi)|, \Psi(\phi) \} = C_{h,\phi}$.  
We also note that $\Psi'$ is non-negative and monotonically decreasing with an asymptote at zero such that $\lim_{z \rightarrow 0} \Psi'(z) = \infty$. Since $c_h < 1$, we have 
$L_{\Psi} = \max_{z \in [c_h \phi, \phi)} \Psi'(z) = \Psi'(c_h \phi)$. Moreover, $\Psi''$ is non-positive and monotonically increasing with $\lim_{z \rightarrow 0} \Psi''(z) = -\infty$ and $\lim_{z \rightarrow \infty} \Psi''(z) = 0$, so that $L_{\Psi'} = \max_{z \in [c_h \phi, \phi)} |\Psi''(z)| = |\Psi''(c_h \phi)|$.

\end{bew}

\begin{lemma}\label{lem:contraction}(Contraction principle) \cite[][Theorem A.3]{vandegeer2008} \\ 
Let $z_1, \ldots, z_n$ be non-random elements of some space $\mc{Z}$, and let 
$\mc{F}$ be a class of real-valued functions on $\mc{Z}$. Consider Lipschitz functions
$\gamma_i: \R \rightarrow \R$, i.e., $|\gamma_i(s) - \gamma_i(\wt{s})| \leq |s-\wt{s}|$ $s,\wt{s} \in \R$, $1 \leq i \leq n$. Let $\eps_1,\ldots,\eps_n$ be i.i.d.~Rademacher variables. Then for any function $f^*: \mc{Z} \rightarrow \R$, we have 
\begin{align*}
\E\left[\sup_{f \in \mc{F}} \left|\su \eps_i \{ \gamma_i(f(z_i)) - \gamma_i(f^*(z_i))\} \right| \right] \leq 2 \E\left[\sup_{f \in \mc{F}} \left|\su \eps_i \{ f(z_i) - f^*(z_i)\} \right|.  \right]
\end{align*}

\end{lemma}

\begin{lemma}\label{lem:Bousquet}(Concentration inequality for empirical processes) \cite[][Corollary 16.1]{vandeGeerSaintFlour}\\ 
Let $\{ Z_i \}_{i = 1}^n$ be independent random variables with range $\mc{Z}$, and let $\mc{F}$ be a class of real-valued functions on $\mc{Z}$ such that $\nnorm{f}_{\infty} < K < \infty$ and 
$\E[f(Z_i)] = 0$, $1 \leq i \leq n$, for all $f \in \mc{F}$. Suppose further that 
$\frac{1}{n} \su \sup_{f \in \mc{F}} \E[f^2(Z_i)]\leq R^2 < \infty$, and let $\M{Z} = \big| \sup_{f \in \mc{F}} \frac{1}{n} \su f(Z_i) \big|$. Then for all $t > 0$
\begin{equation*}
\p \left(\M{Z} \geq 2 \E[\M{Z}] + R \sqrt{\frac{t}{n}} + \frac{4tK}{3n} \right) \leq \exp(-t).
\end{equation*}
\end{lemma}

\end{theo}

\end{document}